\documentclass[final]{agujournal2019}
\usepackage{url} 
\usepackage{lineno}
\usepackage[inline]{trackchanges} 
\usepackage{multirow}
\usepackage{soul}

\draftfalse

\usepackage{subcaption}

\newcommand{\jt}[1]{#1} 
\newcommand{\ue}[1]{#1} 

\journalname{JGR: Atmospheres}

\begin{document}

\title{Calculating radio emissions of positive streamer phenomena using 3D simulations}

\authors{
  Hemaditya Malla\affil{1},
  Yihao Guo\affil{2},
  Brian M.~Hare\affil{3,4},
  Steven Cummer\affil{5},
  Alejandro Malag\'{o}n-Romero\affil{1},
  Ute Ebert\affil{1,2},
  Sander Nijdam\affil{2},
  Jannis Teunissen\affil{1}
}

\affiliation{1}{Centrum Wiskunde \& Informatica (CWI), Amsterdam, The Netherlands}
\affiliation{2}{Department of Applied Physics, Eindhoven University of Technology, The Netherlands}
\affiliation{3}{ASTRON, Dutch Institute for radio astronomy, Dwingeloo}
\affiliation{4}{Kapteyn Astronomical Institute, University of Groningen, Groningen, The Netherlands}
\affiliation{5}{Electrical and Computer Engineering Department, Duke University, Durham, NC, USA}

\correspondingauthor{J.~Teunissen}{jannis.teunissen@cwi.nl}

\begin{keypoints}
\item Streamer branching does not lead to significant radio emission in our simulations and lab experiments.
\item Stochastic fluctuations during streamer propagation increase radio emission at frequencies of 100 MHz and above.
\item When streamers encounter an already partially ionized region, their properties rapidly change leading to emissions up to several 100 MHz.
\end{keypoints}

\begin{abstract}
  We study radio emissions from positive streamers in air using 3D simulations, from which the radiated electric field is computed by solving Jefimenko's equations.
  The simulations are performed at $0.5 \, \textrm{bar}$ using two photoionization methods: the Helmholtz approximation for a photon density and a Monte Carlo method using discrete photons, with the latter being the most realistic.
  We consider cases with single streamers, streamer branching, streamers interacting with preionization and streamer-streamer encounters.
  We do not observe a strong VHF radio signal during or after branching, which is confirmed by lab experiments.
  This indicates that the current inside a streamer discharge evolves approximately continuously during branching.
  On the other hand, stochastic fluctuations in streamer propagation due to Monte Carlo photoionization lead to more radio emission being emitted at frequencies of 100 MHz and above.
  Another process that leads to such high-frequency emission is the interaction of a streamer with a weakly preionized region, which can be present due to a previous discharge.
  In agreement with previous work, we observe the strongest and highest-frequency emission from streamer encounters.
  The amount of total energy that is radiated seems to depend primarily on the background electric field, and less on the particular streamer evolution.
  Finally, we present approximations for the maximal current along a streamer channel and a fit formula for a streamer's current moment.
\end{abstract}

\section*{Plain Language Summary}

The lightning channels in a thunderstorm are preceded by smaller discharges, so-called streamers, which propagate at velocities of hundreds to thousands of kilometers per second.
We cannot see these streamers from the ground, but we can detect their radio emission.
However, it is currently not fully clear what the main mechanisms are by which streamers produce radio emission.
In this paper we therefore perform 3D computer simulations of streamers under different conditions, from which we compute the resulting radio emissions.
We find that streamer branching (the splitting of a streamer channel in two new ones) does not lead to a significant radio signal, which we furthermore have confirmed with lab experiments.
We also show that small fluctuations in a streamer's propagation, which should also occur naturally, lead to radio emission at higher frequencies than in the absence of such fluctuations.
Several other cases are also investigated, such as a ``collision'' between two streamers and streamers in different background electric fields.

\section{Introduction}

Streamer discharges determine the initial stages of electric breakdown in air and other gases \cite{nijdam_physics_2020}. In a tropospheric discharge they form streamer bursts that pave the way of lightning leaders, and streamers appear as sprite discharges in the mesosphere. 
Streamers are space-charge dominated, fast growing plasma filaments with an intricate inner structure. Due to their non-equilibrium nature, they efficiently create nitrogen oxides and ozone, and they also can contribute to electron runaway and consecutive further high energy radiation. 

It is difficult to capture the optical light emission from streamers within a thundercloud, because light is scattered multiple times inside clouds.
Besides visible light, streamers and related lightning phenomena emit electromagnetic radiation across a wide frequency range, from radio frequencies (RF) to microwaves \cite{petersen2014microwave} and beyond \cite{rakov2003lightning}.
Since thunderclouds are transparent to radio emissions in the 3\,kHz--300\,GHz range, such emission can help to understand the phenomena taking place inside a thundercloud \cite{hare2018lofar,liu2022lofar}.

\citeA{brook1964radiation} already measured the radiated electric field from lightning and suggested that the emissions from streamer discharges are in the VHF--UHF range, with VHF (very high frequency) corresponding to 30--300~MHz and UHF (ultra high frequency) corresponding to 0.3--3~GHz.
In recent years, several authors have studied radio emissions from streamer discharges using axisymmetric simulations.
\jt{\citeA{Qin_2012b} simulated sprite streamers and determined their radio emission using the approximations presented by \citeA{uman1975electromagnetic}.
  The authors present an estimate for the decay of the emitted spectrum towards higher frequencies, and they discuss scaling to different gas number densities.}
\citeA{shi2016properties} simulated streamer inception from hydro\-meteors in fields above and below the critical field $E_k$, and computed the radiated electromagnetic field.
\citeA{luque2017radio} performed a full electrodynamic simulation of counter-propagating streamers and their ``collision'' at atmospheric pressure at sub-breakdown conditions.
Such collisions were observed to emit radiation in the range from 100~MHz to a few GHz.
\citeA{shi2019vhf} later studied streamer collisions in background electric fields above $E_k$, and \citeA{koile2021radio} studied collisions in fields below and above $E_k$.
In all these studies, it was found that the background electric field was an important factor controlling the magnitude of the radio emissions.
Furthermore, \citeA{garnung2021hf} studied the emissions of streamer collisions when they occur inside sprites, and they demonstrated that such phenomena could be detected by radiotelescopes.

Radio emissions from streamers have recently also been measured experimentally both in the laboratory and from natural lightning.
\citeA{parkevich2022streamer} and \citeA{parkevich2022electromagnetic} performed laboratory experiments to measure the radio emissions from streamer discharges and found that streamer formation processes emit electromagnetic radiation in the MHz-GHz range.
\citeA{gushchin2021nanosecond} performed laboratory experiments by charging a cloud of water droplets to trigger streamer discharges, and measured electromagnetic radiation in the microwave to UHF range.
\jt{The UHF emissions produced by discharges can be relevant for technological applications, e.g., for partial discharge detection in high-voltage equipment, as discussed by~\citeA{Judd_1996}.}

In natural lightning, VHF and higher frequency radio emissions are routinely observed from almost all negative leaders, presumably from negative polarity streamers.  In contrast, VHF radio emissions from positive leaders and thus positive streamers are much weaker and often undetectable.  \citeA{Pu_2021} reported VHF measurements from a strong and very close positive CG leader and found a VHF spectrum that was mostly flat up until 80 MHz above which it dropped steeply in amplitude.  \citeA{Pu_2022} looked at fast positive breakdown (admittedly an extreme case for positive streamers) and measured a flat VHF-UHF spectrum all the way up to almost 400 MHz.  And \citeA{Scholten_2023} looked for VHF emissions from in-cloud positive leaders and found none despite the very high sensitivity of the LOFAR array.  These three scenarios are all different, but they do show that VHF radio emissions from positive streamers are quite variable, and the underlying mechanism of VHF emissions from positive streamers in natural lightning is poorly understood.  



Typical time scales in a discharge scale approximately inversely with the gas number density $N$~\cite{ebert2010review}.
\jt{Since times scale as $1/N$, frequencies scale as $N$.}
The same type of phenomena can therefore result in radio emission at different frequencies depending on their altitude.
For example, the emissions from head-on sprite streamer collisions at an altitude of 70 km occur in a spectral range from a few kHz up to a few hundred kHz~\cite{garnung2021hf}, whereas the corresponding spectral range for streamer collisions at atmospheric pressure is a few hundred MHz up to a few GHz \cite{luque2017radio,shi2019vhf}.

In this paper, we for the first time compute radio emissions from 3D simulations of positive streamers in air, and we calculate the different spatial components of the electric field at a horizontal distance of 1~km from the approximately vertical discharges.
We consider several cases involving single streamers, streamer branching, the interaction of streamers with a preionized channel and a collision between streamers of opposite polarity.
The simulations are performed with a 3D fluid model in sub-breakdown conditions.
We consider two approaches for photoionization: the so-called Helmholtz approximation where photons are approximated as a density and a Monte Carlo method that takes the full discrete nature of the photons into account~\cite{bagheri2019effect}.
We compute the radiated electric field for all these cases by directly solving Jefimenko's equations for the computed charge density and current density.
Another novel aspect of our work is that we use so-called scaleograms, computed using wavelet analysis, to show how the emission spectrum evolves over time.
Simple estimates for the current and the current moment of a streamer discharge \jt{are provided.
Furthermore, we present} laboratory measurements of the discharge current and the emitted radiation during streamer branching events.

The paper is organized as follows.
We first present the simulation model and the methods for computing and analyzing the radiated field in section~\ref{sec:model}.
Results for all the simulation cases are presented and discussed in section~\ref{sec:results}.
Finally, we present experimental results on the effect of branching on the streamer current in \jt{section~\ref{sec:exper-results-poss}}, and we give numerical details on the solution of Jefimenko's equation in~\ref{sec:jefimenko-numerical}. \ue{\ref{sec:far-field} and \ref{sec:mc-photo} discuss the far field approximation of the radio waves and the numerical implementation of discrete photo-ionization.}



\section{Simulation method}\label{sec:model}

We perform full 3D streamer simulations with the classical drift-diffusion-reaction fluid model using the open-source \texttt{afivo-streamer} code \cite{teunissen_simulating_2017, Teunissen_2018_afivo}.
For a comparison of this model against experiments and against particle simulations see \cite{li_comparing_2021} and \cite{Wang_2022}.
We briefly summarize the main aspects of the model below, further details are given e.g.\ by~\citeA{teunissen_simulating_2017} and \citeA{malla2023double}.

\subsection{Model}
\label{sec:model-description}


In the classical drift-diffusion-reaction fluid model the electron density $n_e$ evolves in time as
\begin{equation}
    \label{eq:ddt_e}
	\partial_t n_e = \nabla \cdot (\mu_e \mathbf{E} n_e + D_e \nabla n_e) + S_e + S_{\mathrm{ph}},
\end{equation}
where $\mu_e$ is the electron mobility coefficient, $D_e$ the diffusion coefficient, $\mathbf{E}$ is the electric field, $S_e$ the source (and sink) term of free electrons due to the processes listed in table~\ref{tbl:reaction_table}, and $S_{\mathrm{ph}}$ is the photo-ionization source term.
In this paper, we consider relatively short time scales of up to about $120 \, \textrm{ns}$ at $500 \, \textrm{mbar}$; this pressure corresponds to an altitude of about 5~km according to the 1976 US standard atmosphere.
The motion of ion and neutral species is therefore not taken into account, so that their densities $n_i$ (for $i=1,\ldots,n$) evolve as
\begin{equation}
  \label{eq:ddt_ni}
  \partial_t n_i = S_i,
\end{equation}
where the $S_i$ are the respective source terms due to reactions.
We use the same reactions as \citeA{li_comparing_2021}; they are given in table~\ref{tbl:reaction_table}.

\begin{table*}
  \centering
  \caption{List of reactions included in the model.}
  \begin{tabular}{ll}
    \hline
    Reaction & Rate coefficient \\ \hline
    $\textrm{e} + \textrm{N}_2 \stackrel{k_1}{\longrightarrow} \textrm{e} + \textrm{e} + \textrm{N}_2^+$ (15.60 eV) & $k_1(E/N)$\\
    $\textrm{e} + \textrm{N}_2 \stackrel{k_2}{\longrightarrow} \textrm{e} + \textrm{e} + \textrm{N}_2^+$ (18.80 eV) & $k_2(E/N)$ \\
    $\textrm{e} + \textrm{O}_2 \stackrel{k_3}{\longrightarrow} \textrm{e} + \textrm{e} + \textrm{O}_2^+$ & $k_3(E/N)$ \\
    $\textrm{e} + \textrm{O}_2 + \textrm{O}_2 \stackrel{k_4}{\longrightarrow} \textrm{O}_2^- + \textrm{O}_2$ & $k_4(E/N)$  \\
    $\textrm{e} + \textrm{O}_2 \stackrel{k_5}{\longrightarrow} \textrm{O}^- + \textrm{O}$ 			& $k_5(E/N)$  \\
    $\textrm{e} + \textrm{N}_2 \stackrel{k_6}{\longrightarrow} \textrm{e} + \textrm{N}_2(\textrm{C}^3 \Pi_u)$ 	& $k_6(E/N)$ \\
    $\textrm{N}_2(\textrm{C}^3 \Pi_u) + \textrm{N}_2 \stackrel{k_7}{\longrightarrow} \textrm{N}_2 + \textrm{N}_2$ & $k_7 = 0.13\times10^{-16}\,\textrm{m}^{3}\textrm{s}^{-1}$  \\
    $\textrm{N}_2(\textrm{C}^3 \Pi_u) + \textrm{O}_2 \stackrel{k_8}{\longrightarrow} \textrm{N}_2 + \textrm{O}_2$ & $k_8 = 3.0\times10^{-16}\,\textrm{m}^3\textrm{s}^{-1}$  \\
    $\textrm{N}_2(\textrm{C}^3 \Pi_u)\stackrel{k_9}{\longrightarrow} \textrm{N}_2(\textrm{B}^3 \Pi_g)$ 		& $k_9=1/(42\,\textrm{ns})$ \\
    \hline
  \end{tabular}
  \label{tbl:reaction_table}
\end{table*}

The electron transport coefficients and the electron-neutral reaction rates $k_1$ to $k_6$ are assumed to depend on the reduced local electric field (local field approximation) $E/N$, where $N$ is the gas number density.
These coefficients were computed using BOLSIG-~\cite{hagelaar_solving_2005} using Phelps' cross-section data for N$_2$ and O$_2$~\cite{Phelps_database}.
We compute optical light emission from the second positive system of N$_2$ using the reaction rates given by \citeA{pancheshnyi_development_2005}. They are given in the last three rows in table~\ref{tbl:reaction_table}.

The photoionization source term $S_{\mathrm{ph}}$ is computed using the model of \citeA{zheleznyak1982photoi_english}, using either the continuum approach like \citeA{Bagheri_2018a} or the discrete Monte-Carlo approach as e.g.\ \citeA{bagheri2019effect} and \citeA{Wang_2023}.
The discrete approach is the most realistic, as it takes the quantization of the photons into account and therefore reproduces the stochastic fluctuations in the electron density ahead of a streamer that should physically be there, \jt{as described in~\ref{sec:mc-photo}}.
  These fluctuations are an important trigger for the branching of streamers, and including them in a model can reproduce the branching observed in experiments~\cite{Wang_2023}.
  The Helmholtz approach has the practical advantage that streamers develop non-stochastically, which can make it easier to study how certain conditions affect the streamer's development.
The two photoionization approaches are compared in section~\ref{sec:single_streamer_results}, and the discrete approach is used in section~\ref{sec:branching_streamer_results} to obtain branching streamers.

The electric field $\mathbf{E}$ is calculated in the electrostatic approximation as $\mathbf{E} = -\nabla \phi$.
Here $\phi$ is the electrostatic potential, which is obtained by solving the Poisson equation with a multigrid method~\cite{Teunissen_2018_afivo}
\begin{equation}
	\nabla^2 \phi = -{\rho}/{\epsilon_0},
\end{equation}
where $\rho$ is the charge density and $\epsilon_0$ is the vacuum permittivity.
\jt{For a discussion of the validity of the electrostatic approximation, see section 5.1 of \cite{nijdam_physics_2020}.}

The electric potential is fixed on the bottom and top of the domain, and homogeneous Neumann boundary conditions are used on the other sides which means that the electrical field is parallel to these lateral boundaries.
In this way, a homogeneous background electric field $E_\mathrm{bg}$ is applied.
For species densities, homogeneous Neumann boundary conditions are used on all domain boundaries, but simulations are stopped before the discharges get close to a boundary.

\subsection{Computational domain}
\label{sec:comp-domain}

\begin{figure}
  \centering
  \includegraphics[width=8cm]{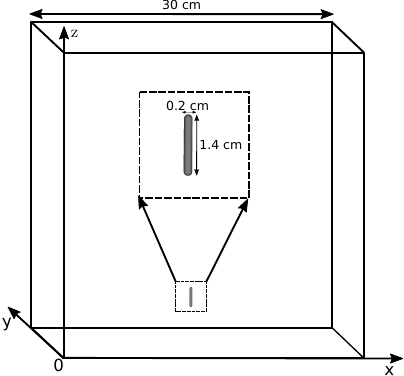}
  \caption{The $(30 \, \textrm{cm})^3$ computational domain, with a zoom into the initial ionization seed in the inlay; such seeds are used to start discharges, see section~\ref{sec:init-cond}.
    Table~\ref{tbl:simulation-cases} lists the locations of the seeds for the different cases considered in this paper. \ue{The maximal ionization density inside the seed is $2.5\cdot 10^{19}$m$^{-3}$.}}
  \label{fig:geometry}
\end{figure}

All simulations are performed in a domain of ($30 \, \textrm{cm}$)$^3$, which is illustrated in figure~\ref{fig:geometry}.
This domain contains dry air (80\% N$_2$, 20\% O$_2$) at a pressure of $500 \, \textrm{mbar}$.
The simulations are performed with adaptive mesh refinement (AMR), as described by \citeA{teunissen_simulating_2017}.

\subsection{Initial conditions}
\label{sec:init-cond}

To start electric discharges we place one or more elongated ionized seeds in the domain.
These seeds are initially electrically neutral.
In the background field they rapidly become polarized, and hence they provide field enhancement and initial electrons, which are both needed for streamer inception.
We use electrically neutral seeds defined by a vertical line segment of 12\,mm and a ``radius'' $r_s = 0.5 \, \mathrm{mm}$ (we remark that the actual radius is larger due to the decay profile).
The density of electrons and O$_2^+$ ions along the seed is given by $n_\mathrm{seed} \cdot f_\mathrm{ss}[(d-r_s)/r_s]$, where $d$ is the distance from the line segment, $n_\mathrm{seed} = 2.5\cdot10^{19}$\,m$^{-3}$ and $f_\mathrm{ss}$ is a smooth step function defined by $f_\mathrm{ss}(x) = 1 - 3 x^2 + 2x^3$ for $0 \leq x \leq 1$, $f_\mathrm{ss}(x) = 1$ for $x < 0$ and $f_\mathrm{ss}(x) = 0$ for $x > 1$.
An example of such a seed is shown in figure~\ref{fig:geometry}.

\subsection{Calculating the radiated electric field}
\label{sec:em_field_equations}

According to a reformulation of the Maxwell equations, the electromagnetic field at a point of observation $\mathbf{r}$ due to an evolving space charge distribution can be computed using Jefimenko's equations~\cite{jefimenko1966electricity,griffiths2005introduction}
\begin{equation}
  \mathbf{E}(\mathbf{r}, t) = \frac{1}{4\pi\epsilon_0}\int \left[ 
    \frac{\rho(\mathbf{r}',t_r)}{\mathcal{R}^2}\hat{\mathbf{\mathcal{R}}}  + \frac{1}{c\mathcal{R}} \partial_t\rho(\mathbf{r}', t_r) \hat{\mathbf{\mathcal{R}}}
    -\frac{1}{c^2\mathcal{R}}\partial_t\mathbf{j}(\mathbf{r}',t_r)\right] \mathrm{d}^3\mathbf{r}',
  \label{eqn:jefimenko_E}
\end{equation}
\begin{equation}
        \mathbf{B}(\mathbf{r}, t) = \frac{\mu_0}{4\pi}\int \left[ 
            \frac{\mathbf{j}(\mathbf{r}', t_r)}{\mathcal{R}^2}
            + \frac{1}{c\mathcal{R}}\partial_t\mathbf{j}(\mathbf{r}',t_r)\right] \times \hat{\mathbf{\mathcal{R}}}~\mathrm{d}^3\mathbf{r}',
        \label{eqn:jefimenko_B}
\end{equation}
where $\mathcal{R}=|\mathbf{r} - \mathbf{r}'|$, $\hat{\mathbf{\mathcal{R}}} = (\mathbf{r} - \mathbf{r}')/|\mathbf{r} - \mathbf{r}'|$, $c$ is the speed of light, $\rho$ is the charge density, $\mathbf{j}$ is the current density, and $t_r$ is the retarded time given by
\begin{equation}
  \label{eq:t-ret}
    t_r = t - \mathcal{R}/c.
\end{equation}


Equations~(\ref{eqn:jefimenko_E}) and (\ref{eqn:jefimenko_B}) are time-dependent generalizations of Coulombs' and Biot-Savart's laws respectively, where the first term in each of these equations corresponds to the static case, and decays like $1/{\cal R}^2$.
Our interest here is in the radiated fields, which correspond to the terms that drop as $1/\mathcal{R}$:
  \begin{equation}
  \mathbf{E}_\mathrm{rad}(\mathbf{r}, t) = \frac{1}{4\pi\epsilon_0}\int \left[ 
    \frac{1}{c\mathcal{R}} \partial_t\rho(\mathbf{r}', t_r) \hat{\mathbf{\mathcal{R}}}
    -\frac{1}{c^2\mathcal{R}}\partial_t\mathbf{j}(\mathbf{r}',t_r)\right] \mathrm{d}^3\mathbf{r}',
  \label{eqn:jefimenko_E_rad}
\end{equation}
\begin{equation}
        \mathbf{B}_\mathrm{rad}(\mathbf{r}, t) = \frac{\mu_0}{4\pi}\int 
            \frac{1}{c\mathcal{R}}\partial_t\mathbf{j}(\mathbf{r}',t_r) \times \hat{\mathbf{\mathcal{R}}}~\mathrm{d}^3\mathbf{r}'.
        \label{eqn:jefimenko_B_rad}
\end{equation}
We numerically compute $\mathbf{E}_\mathrm{rad}(\mathbf{r}_\mathrm{obs}, t)$ at a certain observation location $\mathbf{r}_\mathrm{obs}$ using the output of $\rho$ and $\mathbf{j} = e\mu_e n_e \mathbf{E}$ from our simulations.
The details of this procedure are described in~\ref{sec:jefimenko-numerical}.
Values for $\mathbf{E}_\mathrm{rad}(\mathbf{r}_\mathrm{obs}, t)$ are typically computed every $0.5 \, \textrm{ns}$, which corresponds to a sampling rate of 2\,GHz.
However, for the cases in a background electric field of $10 \, \textrm{kV/cm}$ we use a higher sampling rate of 20\,GHz.

We remark that in previous work considering axisymmetric simulations, authors have typically used the relation 
\begin{equation} \label{eq:E=cB}
E_\mathrm{rad} = c \, B_\mathrm{rad}.
\end{equation}
This relation stems from the far field approximation for electromagnetic waves in vacuum as discussed in more detail in~\ref{sec:far-field}. 
\ue{If the current distribution is approximated by}
a line current $I(z, t)$ in the vertical $z$ direction, the $\phi$ component of the radiated magnetic field from equation~(\ref{eqn:jefimenko_B_rad}) can be computed as
\begin{equation}
  \label{eq:Bphi}
  B_{\mathrm{rad}, \phi}(\mathbf{r}, t) = \frac{\mu_0}{4\pi} \int \frac{\sin{\theta}}{c \mathcal{R}}
  \partial_t I(z, t_r) \, \mathrm{d}z,
\end{equation}
where $\theta$ is the polar angle of the receiver with respect to the source location, see e.g.~\jt{\cite{shi2016properties,Shao_2016}}.
For a far-away observer, ${E}_\mathrm{rad}$ can further be approximated as
\begin{equation}
  \label{eq:Erad-approx}
E_{\rm rad}({\bf r}, t) \approx \frac{\mu_0}{4\pi}\;\frac{\sin{\theta}}{\mathcal{R}}\; \partial_t M(t_r),
\end{equation}
where $M(t)$ is the current moment obtained by integrating $I(z, t)$ over $z$, see e.g.~\cite{Liu_2020}.
Note that the variation in $t_r$ along the source is not taken into account in equation~(\ref{eq:Erad-approx}).
This is a good approximation when $\sin{\theta} \approx 1$ or when the spatial extent of the source is small compared to the shortest wavelength of interest.

In the present paper we work directly with equation~(\ref{eqn:jefimenko_E}) \ue{for arbitrary current and charge density} since we will consider 3D simulations in which currents can flow in all directions, and since we were not sure beforehand what kind of approximations we could make.
In hindsight, we probably could have used equation~(\ref{eq:Erad-approx}).
Details about the numerical evaluation of equation~(\ref{eqn:jefimenko_E}) on an AMR grid are given in~\ref{sec:jefimenko-numerical}.

Note that the effect of the radiated electric field on the discharge itself is not taken into account here, as this effect would typically be very small and because this requires a much more expensive electrodynamic computation of the fields and their influence on the discharge evolution.
As discussed by \citeA{luque2017radio}, there could be a noticeable effect of the radiated field on encounters between streamers of opposite polarities, due to the very rapid change in the current.


\subsection{Time-frequency analysis procedure}\label{sec:cwt_intro}

\begin{figure}
  \centering
  \includegraphics[width=13cm]{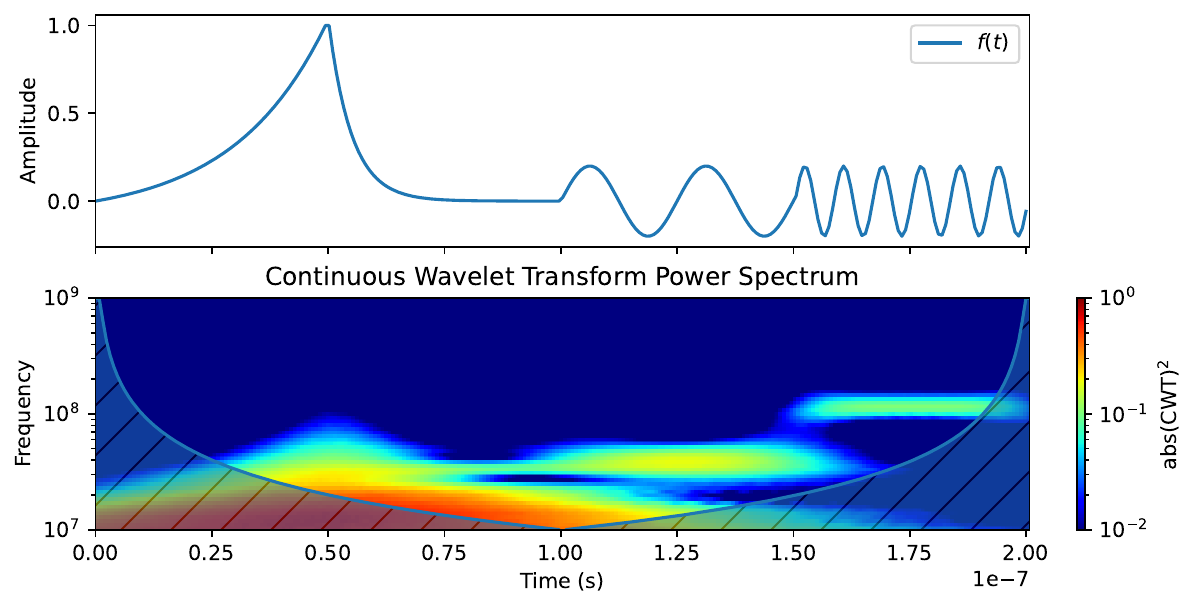}
  \caption{Example of a scaleogram for an artificial signal $f(t)$ consisting of four parts: exponential growth and decay with time constants $\tau = 20 \, \mathrm{ns}$ and $\tau = 5\, \textrm{ns}$, and sine waves with frequencies of $40 \, \textrm{MHz}$ and $120 \, \textrm{MHz}$. The signal was sampled at 256 points. Shown is the so-called power spectrum, in which the color indicates the power in the signal as a function of frequency and time. The regions where values are uncertain due to boundary effects are shaded.
  }
  \label{fig:scaleogram-example}
\end{figure}

We perform a time-frequency analysis of $\mathbf{E}_\mathrm{rad}(\mathbf{r}_\mathrm{obs}, t)$ using a continuous wavelet transform (CWT)~\cite{sejdic2008quantitative, torrence1998practical}, as implemented in the open-source python package \texttt{PyWavelets}~\cite{lee2019pywavelets}.
With this approach, we obtain spectrograms (or more specifically scaleograms, as they are called when doing wavelet analysis) that show how different frequencies are emitted as a function of time.
An example is shown in figure \ref{fig:scaleogram-example}.
For our CWT, we use a Complex-Morlet wavelet~\cite{stephane1999wavelet} with a central frequency of 1~GHz and scale range of [0.1, 100] so that it can resolve frequencies in the range 10~MHz-10~GHz.

The signals we are considering are non-periodic, so some assumption has to be made about the continuation of these signals outside the considered time window.
The regions where values are affected by the boundaries of the signal are indicated in figure~\ref{fig:scaleogram-example}.
We here assume the signal is extended with a constant value, so that if the signal $f(t)$ is given from $t=0$ to $t=T$, we have $f(-T/2 < t < 0) = f(0)$ and $f(T < t < T+T/2) = f(T)$.
In general, $\mathbf{E}_\mathrm{rad}$ at later times will of course not be constant, but this simple assumption has the advantage that $f(t)$ is continuous and that no strong artificial emission occurs due to boundary effects.

\section{Results}\label{sec:results}

In the subsections below we present the simulation cases that are summarized in table~\ref{tbl:simulation-cases}.
The used background electric fields $E_\mathrm{bg}$ range from 6\,kV/cm to 10\,kV/cm, which correspond to about 40\% to 67\% of the critical field at $500 \, \textrm{mbar}$.
Since we fix the electric potential on top and bottom of our computational domain, streamers will experience a higher effective field when they have crossed a significant part of the gap, because the field in their channels will typically be much lower than $E_\mathrm{bg}$.
In all cases, the radiated field is computed at $\mathbf{r}_\mathrm{obs} = (0, 1 \, \mathrm{km}, 0)$, i.e., at 1\,km distance in the $+y$ direction.

\begin{table*}
  \centering
  \caption{List of simulation parameters.
    $E_\mathrm{bg}$ indicates the background field.
    All simulations are performed at 500\,mbar and 300K in a 80\%:20\% N$_2$:O$_2$ gas mixture, so that the critical field is about $E_k \approx 15 \, \mathrm{kV/cm}$.
    The coordinates of the initial seeds (as described in section~\ref{sec:init-cond}) are also given.
    Photoionization is either included with a Helmholtz approximation or with discrete Monte Carlo photons (MC), see section~\ref{sec:model-description}.
  }
  \footnotesize
  \begin{tabular}{lccll}
    \hline
    Description                      & Section                              & $E_\mathrm{bg}/E_k$ & Seed(s) $(x, y, z)$ (cm) & Photoionization \\
    \hline
    Single streamers                 & \ref{sec:single_streamer_results}    & 0.40        & $(15, 15, 10.8)$      & Helmholtz \& MC \\
    Double-headed streamer                 & \ref{sec:double-headed}    & 0.67        & $(15, 15, 10.8)$      & Helmholtz \\
    Guided streamer branching        & \ref{sec:branching_streamer_results} & 0.53        & $(15, 15, 10.8)$    & MC              \\
    Interaction with preionization & \ref{sec:ionized_patch_results}      & 0.40        & $(15, 15, 10.8)$      & Helmholtz       \\
    \multirow{2}*{Streamer collision}               & \multirow{2}*{\ref{sec:collision_results}}          & \multirow{2}*{0.67}       & $(15, 15, 12.0)$,      & \multirow{2}*{Helmholtz}       \\
                                     &                                      &                 & $(15, 15, 16.8)$      &                 \\
    \hline
  \end{tabular}
  \label{tbl:simulation-cases}
\end{table*}

\subsection{Single streamer propagation}\label{sec:single_streamer_results}

\subsubsection{Smooth propagation with Helmholtz photoionization}
\label{sec:smooth-prop-with}

\begin{figure}
  \centering
  \includegraphics[width=13cm]{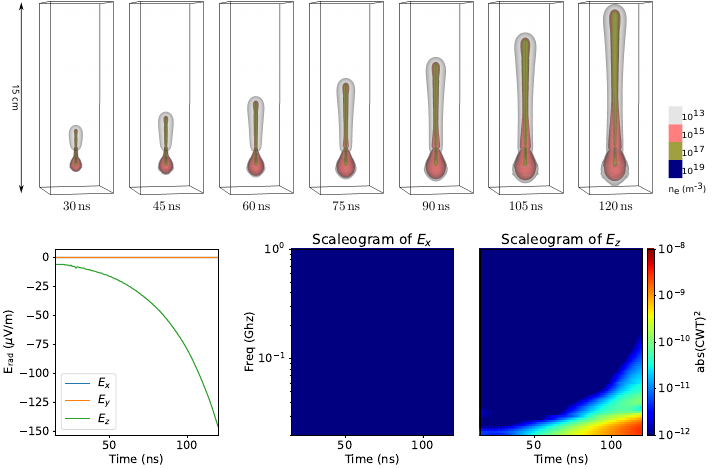}\\
  \caption{Top: contour plots showing the evolution of a single positive streamer with Helmholtz photoionization, propagating in the +z direction in a background electric field of $6 \, \textrm{kV/cm}$.
    Bottom: radiated electric field $\mathbf{E}_\mathrm{rad}$ at $1 \, \mathrm{km}$ distance in the $+y$ direction and corresponding scaleograms.
    The delay due to the travel time of radiation has been subtracted, so that all times are synchronized.
  }
  \label{fig:single_streamer_cont}
\end{figure}

We first simulate a single, non-branching positive streamer starting from a single ionized seed as described in table~\ref{tbl:simulation-cases}, using the Helmholtz photoionization method.
Figure~\ref{fig:single_streamer_cont} shows the time evolution of the electron density, the resulting radiated electric field $\mathbf{E}_\mathrm{rad}$ for an observer at $\mathbf{r}_\mathrm{obs} = (0, 1 \, \mathrm{km}, 0)$, and scaleograms of the different components of $\mathbf{E}_\mathrm{rad}$.
The streamer propagates to a length of 10~cm in 120~ns, with an average velocity of about 10$^6$\,m/s.
As expected, the $x$ and $y$ components of $\mathbf{E}_\mathrm{rad}$ are negligible, and the $z$ component points opposite to the background electric field.
The streamer initially emits weakly in the 10~MHz range, but as its length increases, both the magnitude and the frequency of the emissions increase, as shown in the scaleograms in figure~\ref{fig:single_streamer_cont}.
However, the radiation below 20 MHz remains dominant in amplitude and above 20 MHz the strength falls quickly.  This shape of the radiation spectrum is distinctly different from that reported for strong positive CG leaders \cite{Pu_2021} and for fast positive breakdown \cite{Pu_2022}.  However, VHF radiation that extends only to 20 MHz may be consistent with the lack of detectable radiation from in-cloud positive leaders \cite{Scholten_2023}.

\begin{figure}
  \centering
  \includegraphics[width=13cm]{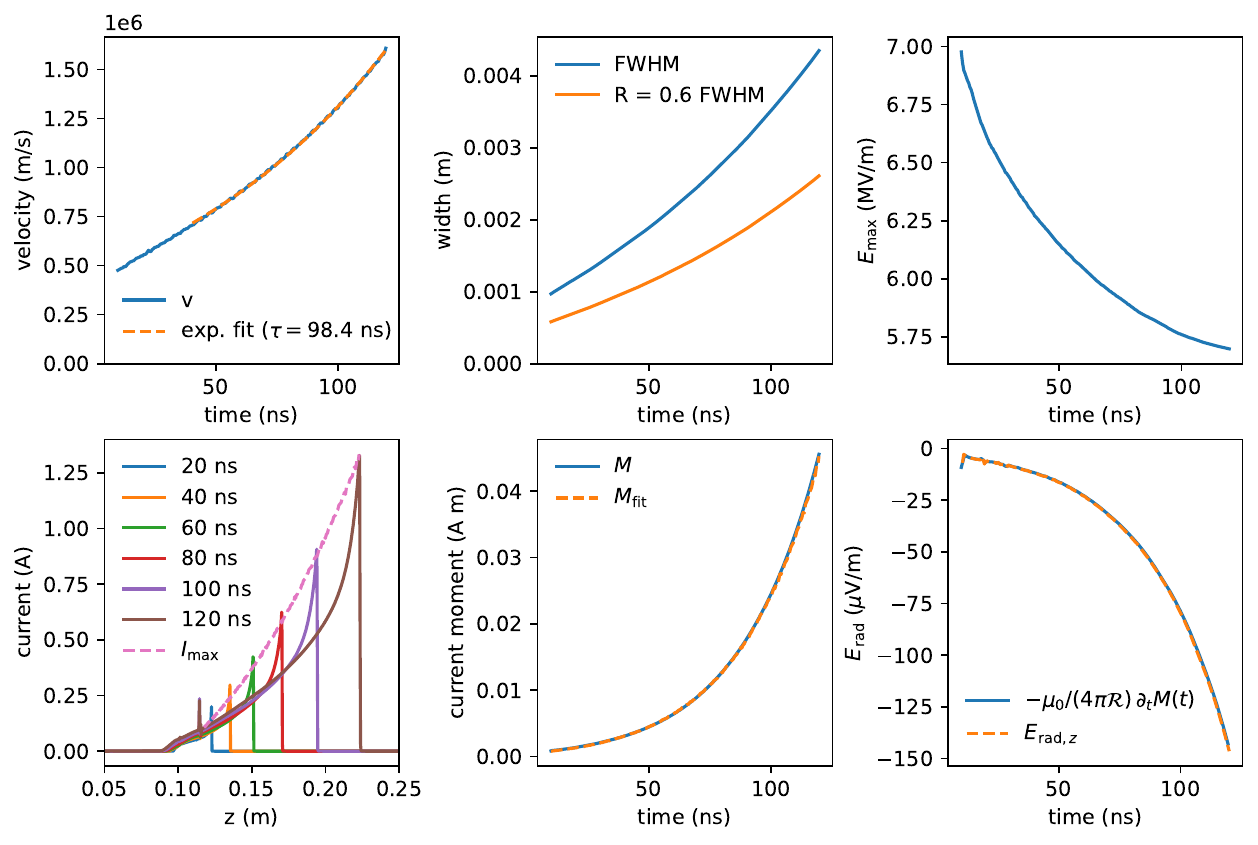}
  \caption{Properties of the single streamer with Helmholtz photoionization.
    Shown on the top row are the velocity, full width half maximum (FWHM) of the optical emission, an estimate for the radius and the maximum electric field at the streamer head. The bottom row shows the current $I(z, t)$ along the $z$ axis, the current moment $M$, and the radiated electric field according to equations~(\ref{eqn:jefimenko_E}) and (\ref{eq:Erad-approx}). The results of equations (\ref{eq:Imax}) and (\ref{eq:M-fit}) are also shown.
  }
  \label{fig:single_streamer_current}
\end{figure}

Figure~\ref{fig:single_streamer_current} shows streamer properties (velocity $v$, radius $R$ and maximum field strength at the streamer head $E_\mathrm{max}$), the current $I$ and current moment $M$, and a comparison between equation~(\ref{eq:Erad-approx}) and equation~(\ref{eqn:jefimenko_E}).
Since this case is essentially axisymmetric, there is almost no difference in the resulting $z$ component of $\mathbf{E}_\mathrm{rad}$.
The radius is here defined as $R = 0.6 \times \, \textrm{FWHM}$, where FWHM is the full width at half maximum of the time-integrated light emission.
The current $I(z, t)$ is computed by integrating the electron conduction current over the $x$ and $y$ directions, and the current moment $M(t)$ is computed by integrating $I(z, t)$ over the $z$ direction.
Both $I$ and $M$ increase smoothly over time.
An estimate for the maximal current $I_\mathrm{max}$ is also shown, given by
\begin{equation}
  \label{eq:Imax}
  I_\mathrm{max} = Q_\mathrm{head} v/R = 2 \pi v R \varepsilon_0 E_\mathrm{max},
\end{equation}
where $Q_\mathrm{head} = 2 \pi R^2 \varepsilon_0 E_\mathrm{max}$ is an approximation for the head charge~\cite{nijdam_physics_2020}.

\jt{Figure~\ref{fig:single_streamer_current} shows that the radiated electric field is to a very good approximation proportional to $\partial_t M(t)$.
  The temporal evolution of the current $I(z, t)$ is also shown, and it is clear that most of the change in $M(t)$ (with $M(t)$ being the area under the curves) is caused by the growth of the streamer head.
  This means that most of the radiation is produced by the streamer head, as was also observed in earlier work, such as that of~\citeA{Qin_2012b}.}

In previous work, \jt{the variation in $M(t)$} has often been characterized as exponential growth~\jt{\cite{Qin_2012b,shi2016properties}}.
Below, we present an estimate for $M(t)$ in terms of streamer properties.
Since $M$ is a spatial integral over $I$, $M$ can be approximated as $M \approx c_1 L I_\mathrm{max}$, where $c_1$ is a constant less than one, $L$ is the length of the channel and $I_\mathrm{max}$ is given by equation~(\ref{eq:Imax}).
Furthermore, on relatively short time scales we can assume that $L \approx c_2 R$ where $c_2$ is a constant greater than one, which states that the radius expands approximately linearly with length.
If the two constants are combined into $c_3 = c_1 c_2$, the result is
\begin{equation}
  \label{eq:M-fit}
  M \approx c_3 R I_\mathrm{max} = (2 \pi c_3 \varepsilon_0) v R^2 E_\mathrm{max},
\end{equation}
and in figure~\ref{fig:single_streamer_current} a curve labeled $M_\mathrm{fit}$ is shown for which $c_3 = 13$.

\jt{\citeA{Qin_2012b} have considered the spectrum produced by a single sprite streamer whose radius and velocity increase approximately exponentially in time.
  If the resulting radiated electric (or magnetic) field also increases exponentially with a time constant $\tau_B$, they argue that the spectrum decays rapidly at higher frequencies with a half-width frequency $f_\mathrm{half} = \sqrt{3}/(2\pi \tau_B)$, where $\tau_B$ depends on the gas number density and the background electric field $E_\mathrm{bg}$.
  We have scaled the tabulated value for $f_\mathrm{half}$ in~\cite{Qin_2012b} at $E_\mathrm{bg} = 0.5 E_k$ to our pressure of $500 \, \textrm{mbar}$, which gave $f_\mathrm{half} \sim 5 \, \mathrm{MHz}$.
  This value agrees quite well with our simulation results; for the power spectrum shown in figure~\ref{fig:single_streamer_cont} we estimate that $f_\mathrm{half} \approx 6 \, \mathrm{MHz}$.
  \citeA{shi2016properties} later on proposed the relation $\tau_B = \tau/3$, where $\tau$ is the time constant of the approximately exponential growth of the velocity.
  We have fitted the velocity shown in figure~\ref{fig:single_streamer_current} for $t > 40 \, \textrm{ns}$ with an exponential function resulting in $\tau \approx 100 \, \textrm{ns}$.
This corresponds to $\tau_B \approx 33 \, \mathrm{ns}$ and thus $f_\mathrm{half} \approx 8 \, \mathrm{MHz}$, also in reasonable agreement with the value of $6 \, \mathrm{MHz}$ estimated from figure~\ref{fig:single_streamer_cont}.}

\subsubsection{Stochastic propagation with Monte Carlo photoionization}
\label{sec:stoch-prop-with}

\begin{figure}
	\centering
	\includegraphics[width=13cm]{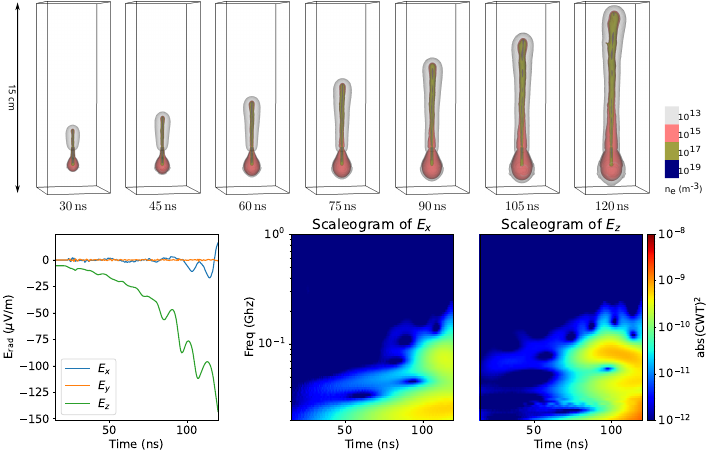}
	\caption{Results for a single streamer with stochastic photoionization, analogous to figure~\ref{fig:single_streamer_cont}.}
	\label{fig:single_streamer_disc}
\end{figure}

We now repeat the single streamer simulation described above using a Monte Carlo photoionization approach, i.e., the photons are modeled as discrete and stochastic.
Figure~\ref{fig:single_streamer_disc} shows the time evolution of the electron density and $\mathbf{E}_\mathrm{rad}$.
The streamer is very similar to the one with Helmholtz photoionization, as it has about the same velocity, radius and electron density profile.
The main difference is that its propagation is now somewhat stochastic, leading to small variations in the direction of propagation and in the streamer radius, and a tiny side branch is visible.

The stochastic growth has several effects on the radiated electric field compared to the smoothly propagating case.
\jt{First}, there is also emission visible in the $x$ direction, since the streamer is no longer axisymmetric.
(Note that $E_{\mathrm{rad}, y} \approx 0$ since the observer is in the $+y$ direction.)
\jt{Second}, there are \jt{oscillations} in $E_{\mathrm{rad}, z}$, which cause the streamer to emit radio emission at higher frequencies during its propagation as shown in the scaleograms.
The period of these oscillations is on the order of 10\,ns, which corresponds to frequencies on the order of $(10\,\mathrm{ns})^{-1} = 100 \, \mathrm{MHz}$.
\jt{We have no direct physical explanation for the period of 10\,ns, but it is about 4-6 times $R/v$, where $R$ is the streamer radius and $v$ its velocity.
The ratio $R/v$ can be thought of as the fastest time scale for a change in streamer head properties.}

The VHF radiation in this simulation extends to approximately 100 MHz with a roughly constant amplitude, with a steep decay above that frequency.  This spectrum is actually quite close to that reported for a strong positive CG \cite{Pu_2021}.  Accordingly it suggests that the presence or lack of detectable VHF emissions from positive leaders (and thus streamers) could be partly driven by the number of streamers: if there are too few, they will not be detectable.


\subsection{Double-headed streamer in a higher background field}
\label{sec:double-headed}

In higher background electric fields streamers tend to propagate and accelerate faster.
Streamers in higher background fields therefore typically radiate more energy, as was found in e.g.~\cite{shi2016properties,luque2017radio,shi2019vhf}.
To compare how both the magnitude and the frequency of the radiation change with the background field, we have performed a simulation using the same conditions as in section~\ref{sec:smooth-prop-with} but with a background electric field of $10 \, \textrm{kV/cm}$ instead of $6 \, \textrm{kV/cm}$.
One other difference is that simulation output was stored more frequently, namely every $0.05 \, \textrm{ns}$.

\begin{figure}
  \centering
  \includegraphics[width=13cm]{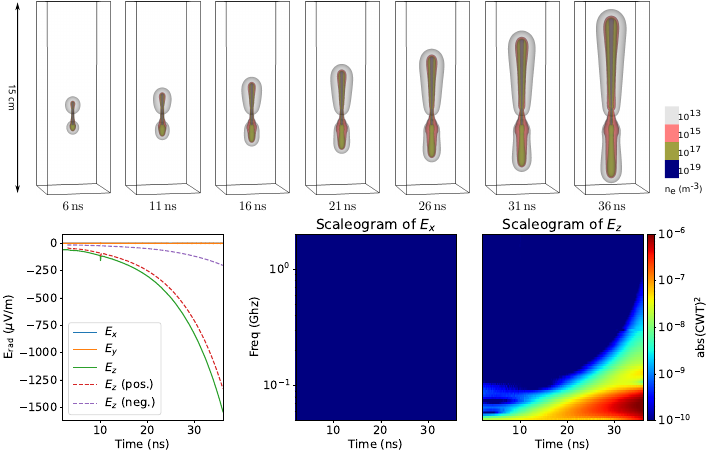}
  \caption{Results for a double-headed streamer developing in a background electric field of $10 \, \textrm{kV/cm}$ with Helmholtz photoionization. Top: electron density contours, bottom: radiated electric field. \jt{The dashed curves show the contributions to from the positive and negative sides of the double-headed streamer.} Note that the scales of frequency and ``power'' abs(CWT)$^2$ in the scaleograms differ from previous plots.}
  \label{fig:voltage_10e5}
\end{figure}

Figure~\ref{fig:voltage_10e5} shows the evolution of the electron density and the radiated electric field.
Due to the higher background field, a double-headed streamer forms.
In about $36 \, \textrm{ns}$, this double-headed streamer grows to a comparable length as the positive streamers in section~\ref{sec:single_streamer_results} did in $120 \, \textrm{ns}$.
We used the Helmholtz photoionization method, so the streamer develops smoothly.
However, relatively smooth propagation would also be expected with Monte Carlo photoionization, because discharges typically develop more smoothly in higher background electric fields (as also observed below in section~\ref{sec:branching_streamer_results}).
The radiated field is considerably stronger than for the case of section~\ref{sec:smooth-prop-with}, with the $z$ component being about an order of magnitude larger.
Furthermore, the scaleograms show that the emission occurs at higher frequencies, as expected.  The higher field produces a faster streamer growth rate and, for this particular set of parameters, strong VHF emissions that are nearly constant in amplitude up to 100 MHz.

\jt{This case contains both a positive and a negative streamer, with the positive streamer propagating upwards in figure~\ref{fig:voltage_10e5} and the negative streamer propagating downwards.
  In the bottom-left plot of figure~\ref{fig:voltage_10e5}, the contributions of the positive and negative sides to the $z$-component of the radiated field are indicated by dashed lines.
  Most of the radio emission is produced by the positive side, since this side propagates about 70\% faster and because its current density $I(z, t)$ is about 4 times higher.
}

\subsection{Branching streamer with stochastic photoionization}
\label{sec:branching_streamer_results}

Most streamer discharges consist of many branched channels, together forming a larger ``streamer tree''.
When a streamer branches, the newly formed channels generally have different radii and velocities than the primary channel, which raises the question whether branching leads to a detectable change in the current through the channel, and therefore to detectable radio emission.
Although it was mentioned by~\citeA{shi2019vhf} that ``branching does not lead to rapid changes in the current moment'', no results were presented to quantify the strength of the effect.

\begin{figure}
	\centering
	\includegraphics[width=13cm]{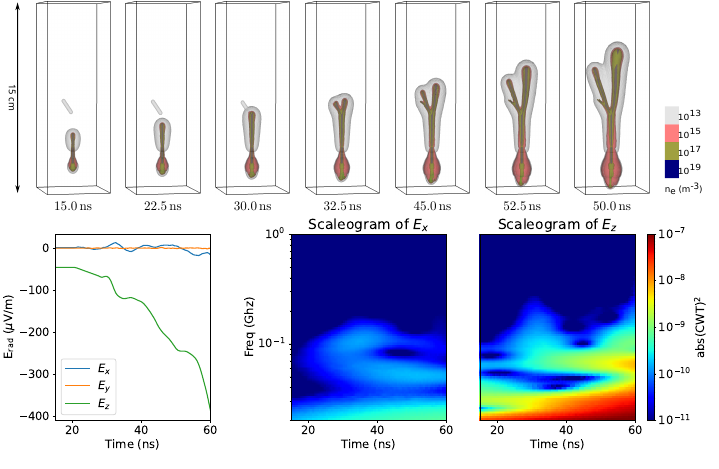}
        \caption{Results for a branching streamer developing in a background electric field of $8 \, \textrm{kV/cm}$ with Monte Carlo photoionization. A small preionized patch is included to trigger the branching, see text. Top: electron density contours, bottom: radiated electric field.}
	\label{fig:branch_11}
\end{figure}

We simulate a branching streamer using the discrete Monte Carlo photoionization model that was also used in section~\ref{sec:stoch-prop-with}, and explained in \ref{sec:mc-photo}.
A slightly higher background electric field of $8 \, \textrm{kV/cm}$ is used, and in order to induce streamer branching early on, we place a weakly ionized patch slightly off-axis as shown in figure~\ref{fig:branch_11}.
This patch contains an equal amount of electrons and O$_2^+$ ions, and it has a degree of ionization of $5 \times 10^{14} \, \textrm{m}^{-3}$.
It is generated using the smoothstep profile as described in section~\ref{sec:init-cond} with $r_s = 0.7 \, \mathrm{mm}$.

Figure~\ref{fig:branch_11} shows that the streamer splits into two somewhat asymmetric branches when it encounters the ionized patch around $t \approx 30 \, \textrm{ns}$.
Although some fluctuations in $E_{\mathrm{rad}, z}$ are visible around this time, these fluctuations are weaker (relatively speaking) than the fluctuations observed without branching in section~\ref{sec:stoch-prop-with}.
During the later propagation, $E_{\mathrm{rad}, z}$ also develops more smoothly than in section~\ref{sec:stoch-prop-with}.
This is caused by the higher background electric field, which generally leads to smoother discharge propagation~\cite{Wang_2023}.
The branches have velocity components in the $+x$ and $-x$ direction, but the radio emission in the $x$ direction is rather weak.

Overall, it seems that streamer branching does not lead to significant variation in the current and thus also not to a significantly different radio emission than the non-branching Monte Carlo streamer shown previously. This is consistent with the earlier statement by~\citeA{shi2019vhf}.
We have also confirmed this finding experimentally, as described in section~\ref{sec:exper-results-poss}.

\subsection{Interaction with ionized patches}\label{sec:ionized_patch_results}

In a large streamer tree, streamer heads can connect to the tails of other streamer channels that have the same polarity.
This phenomenon has been observed in lab experiments, in sprite discharges and in simulations, see e.g.~\cite{Nijdam_2009a,teunissen_simulating_2017}.
We expect that such interactions can lead to a significant change in the current through the channels, and therefore to radio emission.
To test this hypothesis, we simulate the interaction of a single positive streamer with a neutral preionized channel having a much lower degree of ionization, namely $n_0 = 10^{15} \, \mathrm{m}^{-3}$ or $n_0 = 10^{16} \, \mathrm{m}^{-3}$.
Such preionization could correspond to an older streamer, in which most electrons have been lost due to attachment and recombination.

\begin{figure}
	\centering
	\includegraphics[width=13cm]{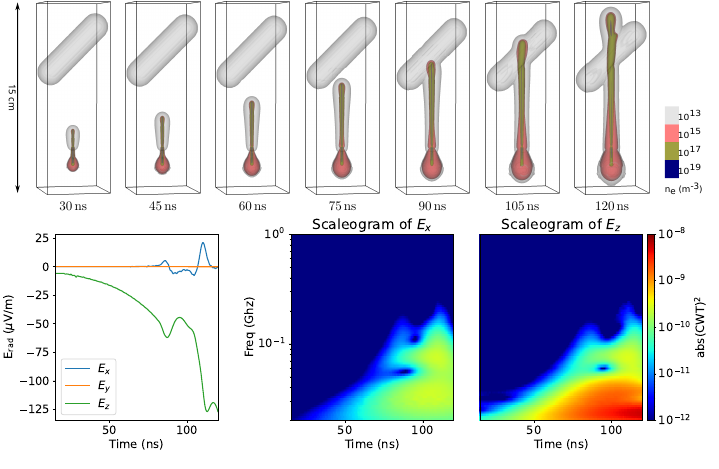}
	\caption{Results for a streamer interacting with a preionized channel in which the electron density is $n_0 = 10^{15} \, \textrm{m}^{-3}$, in a background electric field of $6 \, \textrm{kV/cm}$. Top: electron density contours, bottom: radiated electric field.}
	\label{fig:channel_1e15}
\end{figure}

The preionization is modeled as a cylindrical seed of length $\approx$ 5~cm, using the smoothstep profile described in section~\ref{sec:init-cond} with $r_s = 5 \, \mathrm{mm}$.
The seed is placed at an angle of 45$^\circ$ with the x-axis (see figure~\ref{fig:channel_1e15}).
All other conditions are the same as in section~\ref{sec:smooth-prop-with}, and we now use Helmholtz photoionation.

\begin{figure}
	\centering
	\includegraphics[width=13cm]{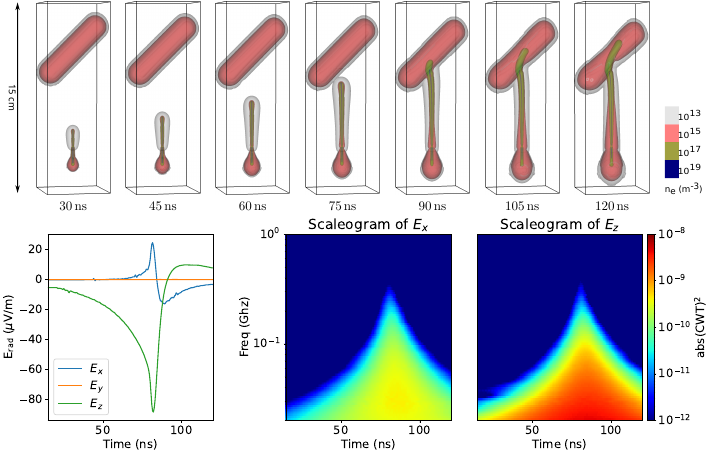}
	\caption{Results analogous to figure~\ref{fig:channel_1e15}, but with a higher electron density in the channel of $n_0 = 10^{16} \, \textrm{m}^{-3}$. The streamer is now strongly deviated by the channel.}
	\label{fig:channel_1e16}
\end{figure}

Simulation results are shown in figures~\ref{fig:channel_1e15} and~\ref{fig:channel_1e16}.
In both cases, the positive streamer reaches the preionized channel at approximately $t=90\,\mathrm{ns}$.
For $n_0 = 10^{15} \, \mathrm{m}^{-3}$, the streamer temporarily slows down as it hits the preionization, leading to a reduction in $E_{\mathrm{rad}, z}$, but it later continues its vertical propagation.
The streamer first slightly deviates towards the patch in the $-x$ direction, and the initial phase of a small side branch is visible in the $+x$ direction.
This development causes a modest signal to be visible in $E_{\mathrm{rad}, x}$.

With $n_0 = 10^{16} \, \mathrm{m}^{-3}$ the preionization has a much stronger effect.
The main streamer is now ``guided'' to propagate inside the preionization~\cite{Nijdam_2016a}.
This leads to an abrupt change in the $z$ component of the current, resulting in a sharp peak in $E_{\mathrm{rad}, z}$ and a sign inversion.
Since the streamer propagates at a roughly 45$^\circ$ angle with respect to the $z$ and $x$ axis,  $E_{\mathrm{rad}, z}$ and $E_{\mathrm{rad}, x}$ are of comparable magnitude during the guided propagation.
The rapid variation in streamer properties and direction generates high-frequency radio emission.
In the scaleograms, signals up to $200-300 \, \textrm{MHz}$ are visible.

\subsection{Streamer collision}\label{sec:collision_results}

\begin{figure}
  \centering
  \includegraphics[width=13cm]{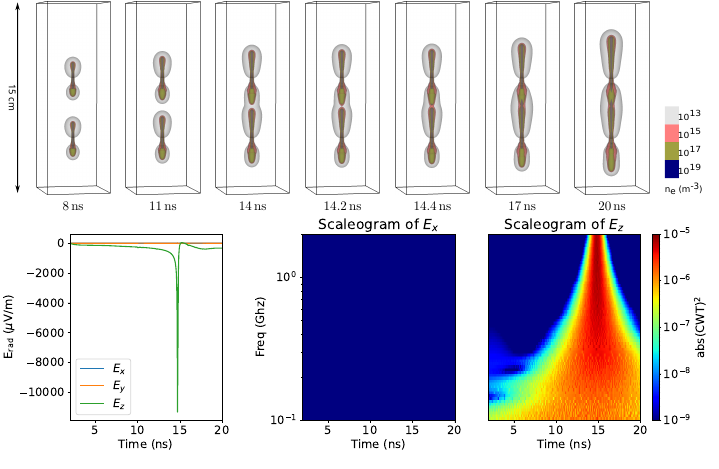}
  \caption{Results for an encounter between a positive streamer propagating upwards and a negative streamer propagating downwards, in a background electric field of $10 \, \textrm{kV/cm}$. Note that two double-headed streamers form. Top: electron density contours, bottom: radiated electric field.}
  \label{fig:collision}
\end{figure}

Encounters or ``collisions'' between opposite polarity streamers have frequently been studied in the literature, see e.g.~\cite{koile2021radio,garnung2021hf,luque2017radio}.
We present one such case here, so that the magnitude of the radiated field can be compared with the other cases that we have studied.
As an initial condition, we use two initial seeds as described in section~\ref{sec:init-cond} and table~\ref{tbl:simulation-cases}.
We apply a higher background electric field of $10 \, \textrm{kV/cm}$ so that two double headed streamers form.
We also store simulation output more frequently, namely every $0.05 \, \textrm{ns}$, so that we can resolve the collision dynamics.
Figure~\ref{fig:collision} shows how a positive and negative streamer connect around $t = 14 \, \textrm{ns}$.
In agreement with previous work, a sharp peak is visible in the $z$ component of the radiated electric field at the moment of collision.
At this time, the emission spectrum extends to multiple GHz, since the collision takes place on sub-nanosecond time scales.
The peak in $E_\mathrm{rad}$ has a significantly larger amplitude than it did for the other cases considered in this paper, but the emission lasts for a much shorter time.

\subsection{Radiated energy}\label{sec:radiated_energy}

To compare how much energy is radiated in these different cases, we have computed the time integral of the Poynting flux of radiated energy
\begin{equation}
  \label{eq:poynting-integral}
  Q_\mathrm{rad} ({\bf r}_\mathrm{obs})
  = \int_{t_0}^{t_1} \frac{{\bf E}\times{\bf B}}{\mu_0}\cdot\hat{\bf r}_\mathrm{obs} \,\mathrm{d}t
  = \int_{t_0}^{t_1} c \, \varepsilon_0 \, \left|{\bf E} ({\bf r}_\mathrm{obs})\right|^2 \,\mathrm{d}t\,
\end{equation}
at the observation location $\mathbf{r}_\mathrm{obs} = (0, 1 \, \mathrm{km}, 0)$, where we have used the identity (\ref{eq:C1}) from \ref{sec:far-field} that is valid for planar, but not necessarily monochromatic waves. 
$Q_{rad}$ has units of energy per unit area and $t_0$ and $t_1$ correspond to the time range shown in the respective plots of $\mathbf{E}_\mathrm{rad}$.
In table~\ref{tbl:energy-radiated}, $Q_\mathrm{rad}$ is given for each case.
The amount of radiated energy seems to primarily depend on the background electric field. Considerably more energy is radiated for the double-headed streamer case and the streamer collision case, which both took place in a background electric field of $10 \, \textrm{kV/cm}$.
For comparison, we have also listed the total energy dissipated in each discharge in table~\ref{tbl:energy-radiated}, which was computed by integrating the Joule heating term over space and time:
\begin{equation}
  \label{eq:deposited-power}
  \varepsilon_\mathrm{tot} = \int_{t_0}^{t_1} \left[\int_V \mathbf{j} \cdot \mathbf{E} \, \mathrm{d}^3{\bf r}\right] \mathrm{d}t\,. 
\end{equation}
The variation in the deposited energy is rather small, which can be explained by the fact that the discharges grow to comparable sizes in all test cases.
From these results, we can furthermore conclude that the fraction of energy that is radiated is significantly larger in a higher background electric field, with the streamer collision being the most `efficient' process.

\begin{table*}
  \centering
  \caption{Total radiated energy per unit area for each case, at a distance of $1 \, \mathrm{km}$ in the $+y$ direction. The total energy deposited in the discharge and the simulation end time are also listed.}
  \begin{tabular}{lcccc}
    \hline
    Description                                     & $E_\mathrm{bg}$ & $Q_\mathrm{rad} \, (\mathrm{J/m}^2)$ & $\varepsilon_\mathrm{tot} \, (\mathrm{J})$ & $t_\mathrm{end} \, (\mathrm{ns}$) \\
    \hline
    Single streamer - Helmholtz                     & 6\,kV/cm        & $9.3 \times 10^{-19}$ & $6.3\times 10^{-4}$ & 120\\
    Single streamer - Monte Carlo                   & 6\,kV/cm        & $8.4 \times 10^{-19}$ & $6.3\times 10^{-4}$ & 120\\
    Preionized channel $10^{15} \, \textrm{m}^{-3}$ & 6\,kV/cm        & $7.5 \times 10^{-19}$ & $6.1\times 10^{-4}$ & 120\\
    Preionized channel $10^{16} \, \textrm{m}^{-3}$ & 6\,kV/cm        & $2.2 \times 10^{-19}$ & $5.2\times 10^{-4}$ & 120\\
    Guided streamer branching                       & 8\,kV/cm        & $3.5 \times 10^{-18}$ & $6.8\times 10^{-4}$ & 60\\
    Single streamer - Helmholtz                     & 10\,kV/cm        & $3.1 \times 10^{-17}$ & $1.1\times 10^{-3}$ & 36\\
    Streamer collision                              & 10\,kV/cm       & $3.4 \times 10^{-17}$ & $4.8\times 10^{-4}$ & 20\\
    \hline
  \end{tabular}
  \label{tbl:energy-radiated}
\end{table*}

\section{\jt{Experimental results on possible radio emission from streamer branching}}
\label{sec:exper-results-poss}

\subsection{Experimental methods}
\begin{figure}
	\centering
	\includegraphics[width=13cm]{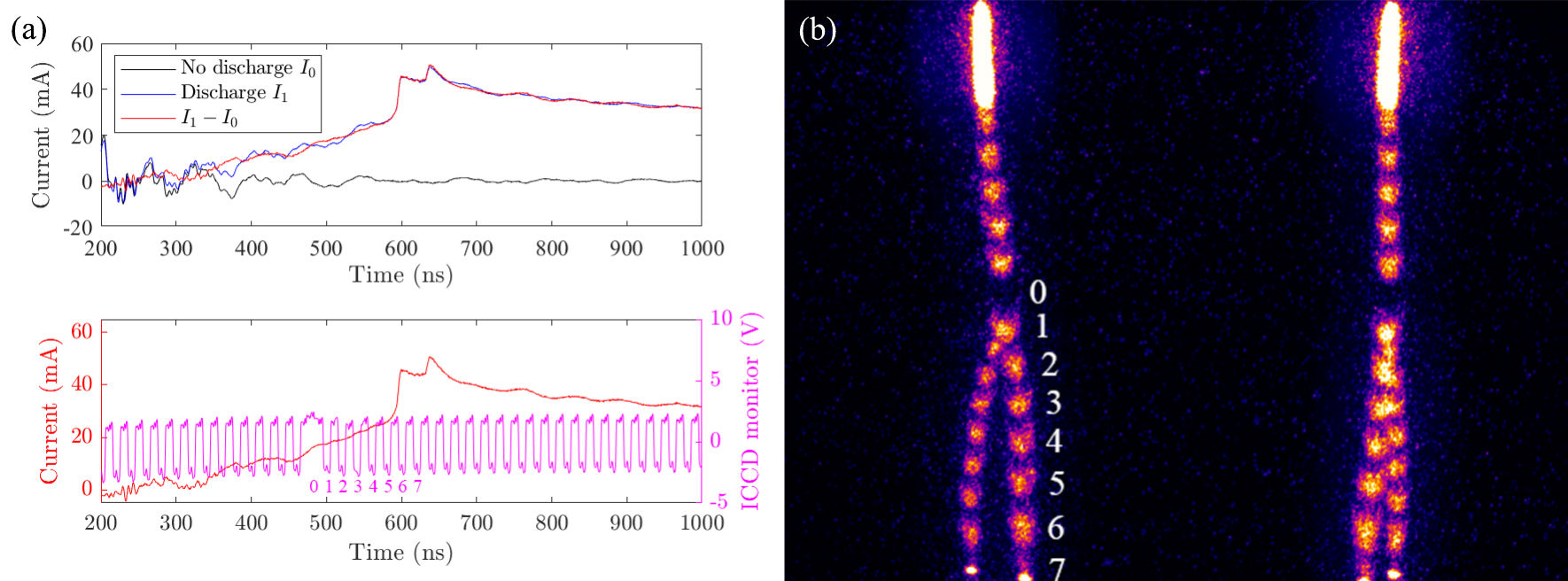}
	\caption{\jt{Experimental results.} (a) Current waveform and camera gating; (b) Corresponding image. \jt{A sharp increase in the current is visible when the streamers cross the gap.}}
    \label{fig:current_gating_image}
\end{figure}

To investigate whether the current shows an apparent peak during streamer branching, we have performed current measurements synchronized with stroboscopic images of streamer discharges in air within a protrusion-to-plane gap. 
The experimental setup we use is similar to the one described by~\citeA{Dijcks2023_imaging}, but with some slight differences. The high voltage was pulsed at a frequency of 20\,Hz, while the CCD operated at 2 frames per second (fps), enabling us to capture one in every ten discharges.
Figure~\ref{fig:current_gating_image}(a) illustrates the typical waveforms of the current and the 50\,MHz camera gating signal used for stroboscopic imaging. 
The current was measured by a 50\,$\Omega$ shunt resistor which connected the bottom plate and the ground. 
Due to the inherent noise in the current measured by the shunt resistor and the presence of damped oscillations from capacitive current in the subsequent waveform, we subtracted the current without discharge ($I_0$) from the measured discharge current ($I_1$). 
Additionally, we applied a Savitzky-Golay filter to smooth the current waveforms. 

In order to establish the relation between image timing and current trace timing, the intensifier gating was set-up to skip one pulse in the 50\,MHz pulse train. 
As a result, the captured image will display a missing dot, corresponding to the skipped pulse, which is shown in figure~\ref{fig:current_gating_image}(b). 
In this way, the timing of every dot on the captured image is known with an accuracy of about 8 ns, which is the open time of the camera gating.

\label{sec:result_antenna}

\subsection{Current change during streamer branching}

\begin{figure}
	\centering
	\includegraphics[width=13cm]{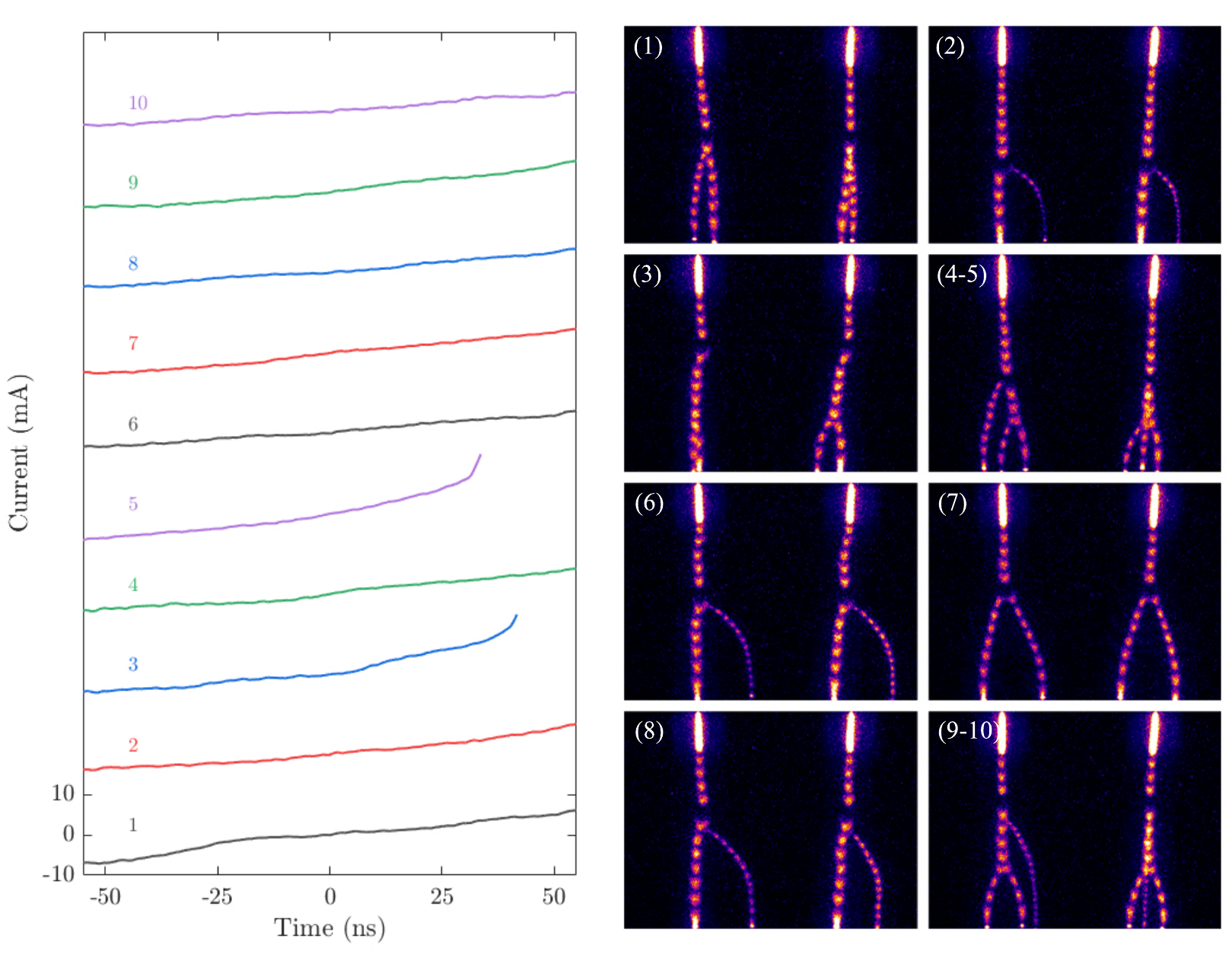}
	\caption{(a) Current waveforms during branching events. 
    The vertical position of each current curve is offset to avoid overlapping and $t = 0$
    corresponds \jt{to} the estimated time of branching; 
    (b) Corresponding discharge images.}
    \label{fig:current_centered}
\end{figure}

In order to visualize the current change during the streamer branching, we shift the current measurements so that the branching deduced from the images occurs at $t = 0$.
Ten examples of current waveforms during branching and the corresponding discharge images are shown in figure~\ref{fig:current_centered}.
For cases that branch more than two times, the corresponding current waveforms are both shown, \textit{e.g.}, waveforms 4-5 and 9-10.
From the figure, we can conclude that no significant change of current can be detected around the streamer branching moment.

\begin{figure}
	\centering
	\includegraphics[width=8cm]{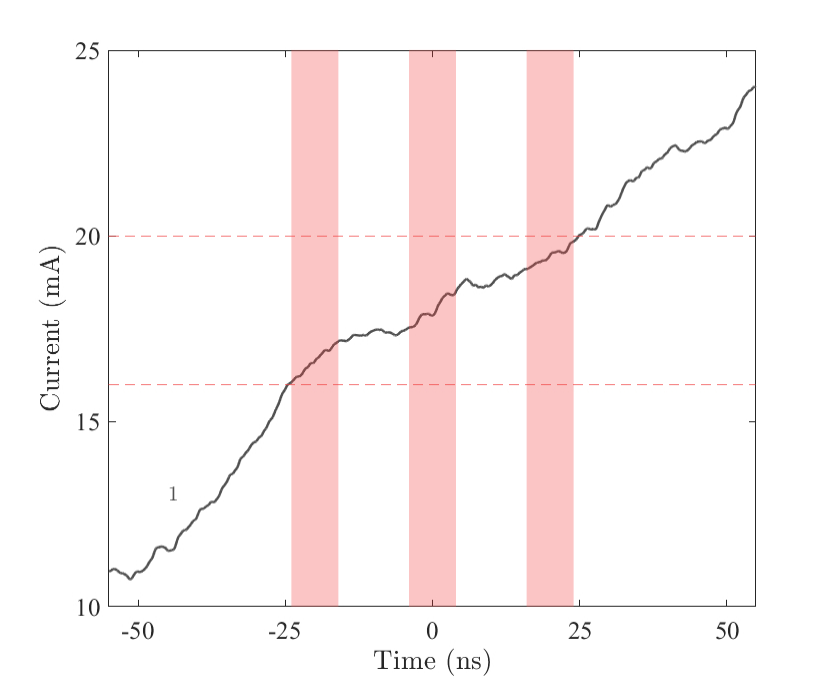}
	\caption{Zoom into current waveform No.~1 from figure~\ref{fig:current_centered}, with the camera gating indicated by a red shade.}
    \label{fig:current_zoom_in}
\end{figure}

\jt{Figure~\ref{fig:current_zoom_in} shows a zoom into current waveform No.~1 from figure~\ref{fig:current_centered}.
  Ripples on the order of 0.2\,mA are present, but these ripples are much smaller than the approximately 4\,mA change in current between two camera gatings (during which the branching event could have taken place).
Since we could not systematically correlate these small ripples to branching events, we associate them with the inherent noise present in our setup.}
From these results, we conclude that the current change during streamer branching is too small to detect with our current equipment.
We should note that the streamers give a very clear current jump and peak effect at the moment they have crossed the gap, which is visible in figure~\ref{fig:current_gating_image}.


\subsection{Radio emission during streamer branching}

Besides current measurements, we have performed radio emission measurements of streamer discharges by using a homemade dipole antenna with a total length of 200\,mm and two commercial antennas (Pulse Electronics NMO150/450/758, a 450\,mm long tri-band monopole antenna and Larid Connectivity EFF6060A3S-10MHF1, a 130 $\times$ 3 $\times$ 0.3\,mm adhesive antenna). 
During the experiment, each antenna was mounted on the quartz window (with a horizontal distance of about 320\,mm to the discharge region) of the vessel horizontally and vertically, respectively, in order to detect the variation of electromagnetic field caused by streamer branching from two orthogonal directions.
The streamer discharge and the background electric field in the vessel are primarily in the vertical direction.
Again, all measurements were synchronized with stroboscopic images to show the timing of the branching events.

\begin{figure}
  \centering
  \includegraphics[width=13cm]{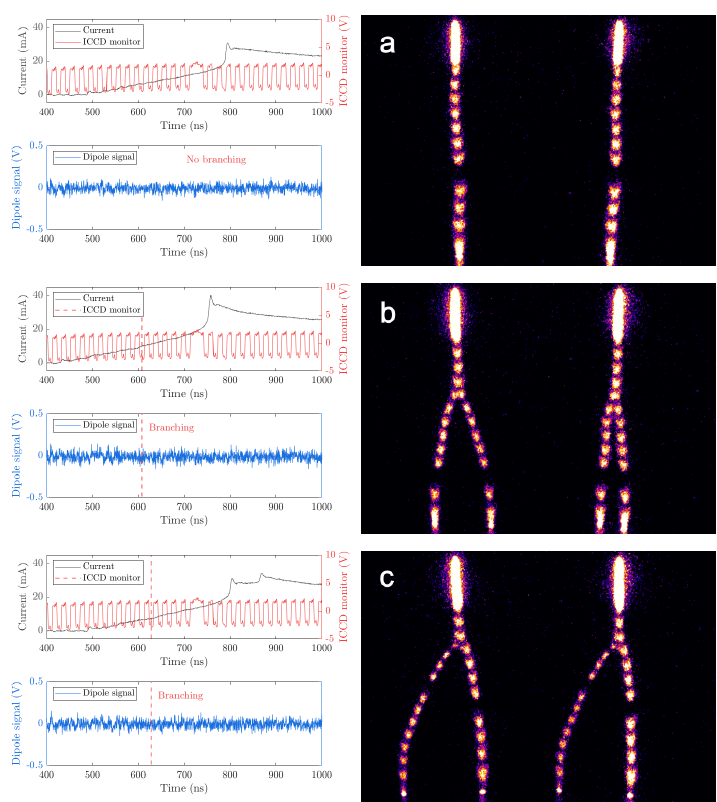}
  \caption{Measurements using a horizontally mounted dipole antenna and the corresponding discharge images for three different discharge morphologies.
    a) Non-branching streamer, b) Branching streamer with equal branches, c) Branching streamer with unequal branches.
    The red dashed lines indicate the approximate timing of the branching event.
  }
  \label{fig:result_antenna}
\end{figure}

As little difference was found between the results from the six different antenna configurations, here we only show the results of the horizontally mounted homemade dipole antenna. 
The response of the antenna was tested by applying standard sinusoidal and pulsed waveforms (with frequency up to 60\,MHz) from a function generator (Keysight 33600A) to the same gas gap. 
The final signals are subtracted by the signal without discharge to eliminate the strong interference produced by the high voltage pulses.
Figure~\ref{fig:result_antenna} shows the current waveforms, the signals of the antenna, and the corresponding discharge images for three different representative cases, namely, a streamer with a single channel, a streamer with two equal branches, and a streamer with two unequal branches. 
For two unequal branches, their velocities, but also their times of arrival are quite distinct, as can be seen by the two peaks at 800 and 880\,ns in figure~\ref{fig:result_antenna}(c).
It should be noted that the real branching event may occur within adjacent camera gating cycles, which can be seen in the discharge image of figure~\ref{fig:result_antenna}(b) and (c). 
Nevertheless, no significant difference can be seen between the signals of the dipole antenna for the three different cases.
Furthermore, no signal is found when streamers cross the gap, even though the current change is much larger than that of branching, implying that the emitted signal is either too small or not in the sensitive range of the antenna. 
Although there are some slight features visible in the signals for other configurations, they cannot be correlated with branching events.

Besides the results shown above, we have also stacked larger numbers
of waveforms like shown in figure~\ref{fig:current_centered}(a) for both
current measurements and antenna output, but did not find any significant signal.
\jt{Therefore, we can conclude from our experiments that the current signal and
the radio emissions produced by streamers do not exhibit any features that we can correlate
with branching events within our experimental limitations.}

\section{\jt{Summary}}

In this paper we have studied radio emissions from vertically moving positive streamers in air at $500 \, \textrm{mbar}$ using 3D simulations.
The electric field radiated by the charge and current densities was computed at 1\,km distance in the horizontal direction by numerically evaluating Jefimenko's equations.
To illustrate the temporal evolution of the emission spectrum, we computed so-called scaleograms using wavelet analysis methods.

We first considered single positive streamers and compared results with two photoionization methods: the Helmholtz approximation and a Monte Carlo method using discrete photons.
With the Helmholtz approximation our simulated streamers grow and accelerate smoothly.
The resulting power spectrum of the radio emissions exhibits a characteristic decay, with the emission shifted to higher frequencies in higher background fields.
With the Monte Carlo approach the overall streamer evolution was highly similar, but small stochastic fluctuations during the streamer propagation led to substantially more radio emission at frequencies of 100 MHz and above.

The Monte Carlo approach is the most realistic approximation of photoionization, as it reproduces the stochastic fluctuations in the electron density ahead of a streamer that should physically be present.
How strongly these fluctuations affect streamer propagation in a thunderstorm will to a large degree depend on the background electric field.
In a higher background electric field, streamer channels generally produce more photoionization per unit length, so that stochastic fluctuations become less important, as shown by~\citeA{Wang_2023}.
Furthermore, at higher altitudes, the amount of photoionization will increase (relatively) due to a reduction in the collisional quenching of photon-emitting molecules.

For the smoothly developing case with Helmholtz photoionization, we presented an approximation for the maximal current along the channel and a fit formula for the current moment as a function of streamer properties.
A double-headed streamer in a higher background electric field was also simulated.
As expected, the faster growth rate of this discharge led to an increase in both the magnitude and the frequency of radio emissions.
\jt{About 80\% of the observed signal came from the positive side of the double-headed streamer, because this side propagated faster and had a higher current density than the negative side.
  We therefore think it is remarkable that the VHF radio emissions from positive leaders and streamers are usually much weaker (and often undetectable) compared to the emissions from negative leaders and streamers.
On the other hand, positive streamers have been observed to produce strong radio emissions during fast positive breakdown events.}


One of the main reasons for performing this study was to understand how much radio emission would be generated by streamer branching, since this process happens very frequently in most streamer discharges.
We simulated branching with the Monte Carlo photoionization method, and we included a small amount of preionization along to trigger the branching process.
We did not observe a strong signal during or after the branching process, which was also confirmed by lab experiments.
This indicates that the current inside a streamer discharges evolves approximately continuously during branching.

Another process that could frequently occur in lightning discharges is the interaction of streamers with some residual ionization, for example from a previous streamer channel.
We simulated such interactions with preionized channels containing an ionization density of $10^{15} \, \textrm{m}^{-3}$ and $10^{16} \, \textrm{m}^{-3}$.
When encountering such pre-ionization, the streamer current and direction can rapidly change, which leads to increased emission at frequencies above 100 MHz.

Finally, we also simulated the encounter between a positive and a negative streamer.
In agreement with previous findings, the channels ``connect'' to each other on a sub-ns time scale, resulting in a sharp peak in the radio emission, with emission at frequencies above a GHz.
This process emitted the highest amount of radiated energy, but only slightly more than the double-headed streamer in the same background electric field.
We therefore conclude that the amount of radiated energy primarily depends on the background electric field and to a lesser extent on the type of streamer process taking place.
The streamer collision process was the most efficient in terms of converting ``discharge energy'' (the integral over $\mathbf{J} \cdot \mathbf{E}$) into radiated energy.

\jt{In this paper, we have studied the radio emission generated by relatively small streamer discharges.
When discharges occur in or above a thundercloud, there would typically be a large number of streamers growing simultaneously in different regions.
The radio emission of such a composite system can be estimated according to the theory presented by \citeA{Liu_2020}, with low-frequency components adding coherently while high-frequency components add incoherently.}


\section*{Data Availability}

\jt{Version 0.3.1 of \texttt{afivo-streamer}~\cite{teunissen_simulating_2017} used for the streamer simulations is preserved at \url{https://github.com/MD-CWI/afivo-streamer} with a GPL-3.0 license.
The data and tools used in this study~\cite{zenodo-radio} are available at Zenodo via \url{https://zenodo.org/doi/10.5281/zenodo.10977532} with a Creative Commons Attribution 4.0 International license.
Specifically, we provide a tool to solve Jefimenko's equations, which converts output of \texttt{afivo-streamer} simulations to the radiated field at a particular location.
The radiated fields obtained with this tool are provided, as well as simulation input files.
The data set also contains the experimentally obtained discharge images and recorded waveforms.
}

\acknowledgments
H.M. was supported by project NWO-17183 ``Plasma for Plants".
B.M.H.~was supported by ERC Grant Agreement No.~101041097.
Y.G.~was supported by the China Scholarship Council (CSC) Grant No.~202006280041. A.M.R.~was supported by a Ram\'{o}n Areces Foundation grant No.~BEVP34A6840.
The simulations in this work were carried out on the Dutch national e-infrastructure with the
support of SURF Co-operative.

\bibliography{bibliography}

\begin{thebibliography}{}

\bibitem [\protect \citeauthoryear {%
Bagheri%
\ \BBA {} Teunissen%
}{%
Bagheri%
\ \BBA {} Teunissen%
}{%
{\protect \APACyear {2019}}%
}]{%
bagheri2019effect}
\APACinsertmetastar {%
bagheri2019effect}%
\begin{APACrefauthors}%
Bagheri, B.%
\BCBT {}\ \BBA {} Teunissen, J.%
\end{APACrefauthors}%
\unskip\
\newblock
\APACrefYearMonthDay{2019}{}{}.
\newblock
{\BBOQ}\APACrefatitle {The effect of the stochasticity of photoionization on 3D
  streamer simulations} {The effect of the stochasticity of photoionization on
  3d streamer simulations}.{\BBCQ}
\newblock
\APACjournalVolNumPages{Plasma Sources Science and Technology}{28}{4}{045013}.
\PrintBackRefs{\CurrentBib}

\bibitem [\protect \citeauthoryear {%
Bagheri%
\ \protect \BOthers {.}}{%
Bagheri%
\ \protect \BOthers {.}}{%
{\protect \APACyear {2018}}%
}]{%
Bagheri_2018a}
\APACinsertmetastar {%
Bagheri_2018a}%
\begin{APACrefauthors}%
Bagheri, B.%
, Teunissen, J.%
, Ebert, U.%
, Becker, M\BPBI M.%
, Chen, S.%
, Ducasse, O.%
\BDBL {}Yousfi, M.%
\end{APACrefauthors}%
\unskip\
\newblock
\APACrefYearMonthDay{2018}{{\APACmonth{09}}}{}.
\newblock
{\BBOQ}\APACrefatitle {Comparison of Six Simulation Codes for Positive
  Streamers in Air} {Comparison of six simulation codes for positive streamers
  in air}.{\BBCQ}
\newblock
\APACjournalVolNumPages{Plasma Sources Science and Technology}{27}{9}{095002}.
\newblock
\begin{APACrefDOI} \doi{10.1088/1361-6595/aad768} \end{APACrefDOI}
\PrintBackRefs{\CurrentBib}

\bibitem [\protect \citeauthoryear {%
Brook%
\ \BBA {} Kitagawa%
}{%
Brook%
\ \BBA {} Kitagawa%
}{%
{\protect \APACyear {1964}}%
}]{%
brook1964radiation}
\APACinsertmetastar {%
brook1964radiation}%
\begin{APACrefauthors}%
Brook, M.%
\BCBT {}\ \BBA {} Kitagawa, N.%
\end{APACrefauthors}%
\unskip\
\newblock
\APACrefYearMonthDay{1964}{}{}.
\newblock
{\BBOQ}\APACrefatitle {Radiation from lightning discharges in the frequency
  range 400 to 1000 Mc/s} {Radiation from lightning discharges in the frequency
  range 400 to 1000 mc/s}.{\BBCQ}
\newblock
\APACjournalVolNumPages{Journal of Geophysical Research}{69}{12}{2431--2434}.
\PrintBackRefs{\CurrentBib}

\bibitem [\protect \citeauthoryear {%
Chanrion%
\ \BBA {} Neubert%
}{%
Chanrion%
\ \BBA {} Neubert%
}{%
{\protect \APACyear {2008}}%
}]{%
chanrion2008pic}
\APACinsertmetastar {%
chanrion2008pic}%
\begin{APACrefauthors}%
Chanrion, O.%
\BCBT {}\ \BBA {} Neubert, T.%
\end{APACrefauthors}%
\unskip\
\newblock
\APACrefYearMonthDay{2008}{}{}.
\newblock
{\BBOQ}\APACrefatitle {A PIC-MCC code for simulation of streamer propagation in
  air} {A pic-mcc code for simulation of streamer propagation in air}.{\BBCQ}
\newblock
\APACjournalVolNumPages{Journal of Computational Physics}{227}{15}{7222--7245}.
\PrintBackRefs{\CurrentBib}

\bibitem [\protect \citeauthoryear {%
Dijcks%
, Leegte%
\BCBL {}\ \BBA {} Nijdam%
}{%
Dijcks%
\ \protect \BOthers {.}}{%
{\protect \APACyear {2023}}%
}]{%
Dijcks2023_imaging}
\APACinsertmetastar {%
Dijcks2023_imaging}%
\begin{APACrefauthors}%
Dijcks, S.%
, Leegte, M\BPBI v\BPBI d.%
\BCBL {}\ \BBA {} Nijdam, S.%
\end{APACrefauthors}%
\unskip\
\newblock
\APACrefYearMonthDay{2023}{{\APACmonth{04}}}{}.
\newblock
{\BBOQ}\APACrefatitle {Imaging and reconstruction of positive streamer
  discharge tree structures} {Imaging and reconstruction of positive streamer
  discharge tree structures}.{\BBCQ}
\newblock
\APACjournalVolNumPages{Plasma Sources Science and Technology}{32}{4}{045004}.
\newblock
\begin{APACrefURL}
  [{2023-07-06}]\url{https://dx.doi.org/10.1088/1361-6595/acc821}
  \end{APACrefURL}
\newblock
\begin{APACrefDOI} \doi{10.1088/1361-6595/acc821} \end{APACrefDOI}
\PrintBackRefs{\CurrentBib}

\bibitem [\protect \citeauthoryear {%
Ebert%
\ \protect \BOthers {.}}{%
Ebert%
\ \protect \BOthers {.}}{%
{\protect \APACyear {2010}}%
}]{%
ebert2010review}
\APACinsertmetastar {%
ebert2010review}%
\begin{APACrefauthors}%
Ebert, U.%
, Nijdam, S.%
, Li, C.%
, Luque, A.%
, Briels, T.%
\BCBL {}\ \BBA {} van Veldhuizen, E.%
\end{APACrefauthors}%
\unskip\
\newblock
\APACrefYearMonthDay{2010}{}{}.
\newblock
{\BBOQ}\APACrefatitle {Review of recent results on streamer discharges and
  discussion of their relevance for sprites and lightning} {Review of recent
  results on streamer discharges and discussion of their relevance for sprites
  and lightning}.{\BBCQ}
\newblock
\APACjournalVolNumPages{Journal of Geophysical Research: Space
  Physics}{115}{A7}{}.
\PrintBackRefs{\CurrentBib}

\bibitem [\protect \citeauthoryear {%
Garnung%
, Celestin%
\BCBL {}\ \BBA {} Farges%
}{%
Garnung%
\ \protect \BOthers {.}}{%
{\protect \APACyear {2021}}%
}]{%
garnung2021hf}
\APACinsertmetastar {%
garnung2021hf}%
\begin{APACrefauthors}%
Garnung, M.%
, Celestin, S.%
\BCBL {}\ \BBA {} Farges, T.%
\end{APACrefauthors}%
\unskip\
\newblock
\APACrefYearMonthDay{2021}{}{}.
\newblock
{\BBOQ}\APACrefatitle {Hf-vhf electromagnetic emissions from collisions of
  sprite streamers} {Hf-vhf electromagnetic emissions from collisions of sprite
  streamers}.{\BBCQ}
\newblock
\APACjournalVolNumPages{Journal of Geophysical Research: Space
  Physics}{126}{6}{e2020JA028824}.
\PrintBackRefs{\CurrentBib}

\bibitem [\protect \citeauthoryear {%
Griffiths%
}{%
Griffiths%
}{%
{\protect \APACyear {2017}}%
}]{%
griffiths2005introduction}
\APACinsertmetastar {%
griffiths2005introduction}%
\begin{APACrefauthors}%
Griffiths, D\BPBI J.%
\end{APACrefauthors}%
\unskip\
\newblock
\APACrefYearMonthDay{2017}{}{}.
\newblock
\APACrefbtitle {Introduction to electrodynamics, 4th Edition.} {Introduction to
  electrodynamics, 4th edition.}
\newblock
\APACaddressPublisher{}{Cambridge University Press}.
\PrintBackRefs{\CurrentBib}

\bibitem [\protect \citeauthoryear {%
Gushchin%
\ \protect \BOthers {.}}{%
Gushchin%
\ \protect \BOthers {.}}{%
{\protect \APACyear {2021}}%
}]{%
gushchin2021nanosecond}
\APACinsertmetastar {%
gushchin2021nanosecond}%
\begin{APACrefauthors}%
Gushchin, M.%
, Korobkov, S.%
, Zudin, I\BPBI Y.%
, Nikolenko, A.%
, Mikryukov, P.%
, Syssoev, V.%
\BDBL {}others%
\end{APACrefauthors}%
\unskip\
\newblock
\APACrefYearMonthDay{2021}{}{}.
\newblock
{\BBOQ}\APACrefatitle {Nanosecond electromagnetic pulses generated by electric
  discharges: Observation with clouds of charged water droplets and
  implications for lightning} {Nanosecond electromagnetic pulses generated by
  electric discharges: Observation with clouds of charged water droplets and
  implications for lightning}.{\BBCQ}
\newblock
\APACjournalVolNumPages{Geophysical Research Letters}{48}{7}{e2020GL092108}.
\PrintBackRefs{\CurrentBib}

\bibitem [\protect \citeauthoryear {%
Hagelaar%
\ \BBA {} Pitchford%
}{%
Hagelaar%
\ \BBA {} Pitchford%
}{%
{\protect \APACyear {2005}}%
}]{%
hagelaar_solving_2005}
\APACinsertmetastar {%
hagelaar_solving_2005}%
\begin{APACrefauthors}%
Hagelaar, G\BPBI J\BPBI M.%
\BCBT {}\ \BBA {} Pitchford, L\BPBI C.%
\end{APACrefauthors}%
\unskip\
\newblock
\APACrefYearMonthDay{2005}{{\APACmonth{11}}}{}.
\newblock
{\BBOQ}\APACrefatitle {Solving the {{Boltzmann}} Equation to Obtain Electron
  Transport Coefficients and Rate Coefficients for Fluid Models} {Solving the
  {{Boltzmann}} equation to obtain electron transport coefficients and rate
  coefficients for fluid models}.{\BBCQ}
\newblock
\APACjournalVolNumPages{Plasma Sources Science and
  Technology}{14}{4}{722--733}.
\newblock
\begin{APACrefDOI} \doi{10.1088/0963-0252/14/4/011} \end{APACrefDOI}
\PrintBackRefs{\CurrentBib}

\bibitem [\protect \citeauthoryear {%
Hare%
\ \protect \BOthers {.}}{%
Hare%
\ \protect \BOthers {.}}{%
{\protect \APACyear {2018}}%
}]{%
hare2018lofar}
\APACinsertmetastar {%
hare2018lofar}%
\begin{APACrefauthors}%
Hare, B.%
, Scholten, O.%
, Bonardi, A.%
, Buitink, S.%
, Corstanje, A.%
, Ebert, U.%
\BDBL {}others%
\end{APACrefauthors}%
\unskip\
\newblock
\APACrefYearMonthDay{2018}{}{}.
\newblock
{\BBOQ}\APACrefatitle {LOFAR lightning imaging: Mapping lightning with
  nanosecond precision} {Lofar lightning imaging: Mapping lightning with
  nanosecond precision}.{\BBCQ}
\newblock
\APACjournalVolNumPages{Journal of Geophysical Research:
  Atmospheres}{123}{5}{2861--2876}.
\PrintBackRefs{\CurrentBib}

\bibitem [\protect \citeauthoryear {%
Jefimenko%
}{%
Jefimenko%
}{%
{\protect \APACyear {1966}}%
}]{%
jefimenko1966electricity}
\APACinsertmetastar {%
jefimenko1966electricity}%
\begin{APACrefauthors}%
Jefimenko, O\BPBI D.%
\end{APACrefauthors}%
\unskip\
\newblock
\APACrefYear{1966}.
\newblock
\APACrefbtitle {Electricity and Magnetism: An Introduction to the Theory of
  Electric and Magnetic Fields} {Electricity and magnetism: An introduction to
  the theory of electric and magnetic fields}.
\newblock
\APACaddressPublisher{}{Appleton-Century-Crofts}.
\PrintBackRefs{\CurrentBib}

\bibitem [\protect \citeauthoryear {%
Judd%
, Farish%
\BCBL {}\ \BBA {} Hampton%
}{%
Judd%
\ \protect \BOthers {.}}{%
{\protect \APACyear {1996}}%
}]{%
Judd_1996}
\APACinsertmetastar {%
Judd_1996}%
\begin{APACrefauthors}%
Judd, M.%
, Farish, O.%
\BCBL {}\ \BBA {} Hampton, B.%
\end{APACrefauthors}%
\unskip\
\newblock
\APACrefYearMonthDay{1996}{{\APACmonth{04}}}{}.
\newblock
{\BBOQ}\APACrefatitle {The Excitation of {{UHF}} Signals by Partial Discharges
  in {{GIS}}} {The excitation of {{UHF}} signals by partial discharges in
  {{GIS}}}.{\BBCQ}
\newblock
\APACjournalVolNumPages{IEEE Transactions on Dielectrics and Electrical
  Insulation}{3}{2}{213--228}.
\newblock
\begin{APACrefDOI} \doi{10.1109/94.486773} \end{APACrefDOI}
\PrintBackRefs{\CurrentBib}

\bibitem [\protect \citeauthoryear {%
Koile%
, Liu%
\BCBL {}\ \BBA {} Dwyer%
}{%
Koile%
\ \protect \BOthers {.}}{%
{\protect \APACyear {2021}}%
}]{%
koile2021radio}
\APACinsertmetastar {%
koile2021radio}%
\begin{APACrefauthors}%
Koile, J.%
, Liu, N.%
\BCBL {}\ \BBA {} Dwyer, J.%
\end{APACrefauthors}%
\unskip\
\newblock
\APACrefYearMonthDay{2021}{}{}.
\newblock
{\BBOQ}\APACrefatitle {Radio frequency emissions from streamer collisions in
  subbreakdown fields} {Radio frequency emissions from streamer collisions in
  subbreakdown fields}.{\BBCQ}
\newblock
\APACjournalVolNumPages{Geophysical Research Letters}{48}{24}{e2021GL096214}.
\PrintBackRefs{\CurrentBib}

\bibitem [\protect \citeauthoryear {%
Lee%
, Gommers%
, Waselewski%
, Wohlfahrt%
\BCBL {}\ \BBA {} O'Leary%
}{%
Lee%
\ \protect \BOthers {.}}{%
{\protect \APACyear {2019}}%
}]{%
lee2019pywavelets}
\APACinsertmetastar {%
lee2019pywavelets}%
\begin{APACrefauthors}%
Lee, G.%
, Gommers, R.%
, Waselewski, F.%
, Wohlfahrt, K.%
\BCBL {}\ \BBA {} O'Leary, A.%
\end{APACrefauthors}%
\unskip\
\newblock
\APACrefYearMonthDay{2019}{}{}.
\newblock
{\BBOQ}\APACrefatitle {PyWavelets: A Python package for wavelet analysis}
  {Pywavelets: A python package for wavelet analysis}.{\BBCQ}
\newblock
\APACjournalVolNumPages{Journal of Open Source Software}{4}{36}{1237}.
\PrintBackRefs{\CurrentBib}

\bibitem [\protect \citeauthoryear {%
Li%
\ \protect \BOthers {.}}{%
Li%
\ \protect \BOthers {.}}{%
{\protect \APACyear {2021}}%
}]{%
li_comparing_2021}
\APACinsertmetastar {%
li_comparing_2021}%
\begin{APACrefauthors}%
Li, X.%
, Dijcks, S.%
, Nijdam, S.%
, Sun, A.%
, Ebert, U.%
\BCBL {}\ \BBA {} Teunissen, J.%
\end{APACrefauthors}%
\unskip\
\newblock
\APACrefYearMonthDay{2021}{sep}{}.
\newblock
{\BBOQ}\APACrefatitle {Comparing simulations and experiments of positive
  streamers in air: steps toward model validation} {Comparing simulations and
  experiments of positive streamers in air: steps toward model
  validation}.{\BBCQ}
\newblock
\APACjournalVolNumPages{Plasma Sources Science and Technology}{30}{9}{095002}.
\newblock
\begin{APACrefURL} \url{https://doi.org/10.1088/1361-6595/ac1b36}
  \end{APACrefURL}
\newblock
\begin{APACrefDOI} \doi{10.1088/1361-6595/ac1b36} \end{APACrefDOI}
\PrintBackRefs{\CurrentBib}

\bibitem [\protect \citeauthoryear {%
N.~Liu%
, Dwyer%
\BCBL {}\ \BBA {} Tilles%
}{%
N.~Liu%
\ \protect \BOthers {.}}{%
{\protect \APACyear {2020}}%
}]{%
Liu_2020}
\APACinsertmetastar {%
Liu_2020}%
\begin{APACrefauthors}%
Liu, N.%
, Dwyer, J\BPBI R.%
\BCBL {}\ \BBA {} Tilles, J\BPBI N.%
\end{APACrefauthors}%
\unskip\
\newblock
\APACrefYearMonthDay{2020}{{\APACmonth{07}}}{}.
\newblock
{\BBOQ}\APACrefatitle {Electromagnetic {{Radiation Spectrum}} of a {{Composite
  System}}} {Electromagnetic {{Radiation Spectrum}} of a {{Composite
  System}}}.{\BBCQ}
\newblock
\APACjournalVolNumPages{Physical Review Letters}{125}{2}{025101}.
\newblock
\begin{APACrefDOI} \doi{10.1103/PhysRevLett.125.025101} \end{APACrefDOI}
\PrintBackRefs{\CurrentBib}

\bibitem [\protect \citeauthoryear {%
N\BPBI Y.~Liu%
\ \protect \BOthers {.}}{%
N\BPBI Y.~Liu%
\ \protect \BOthers {.}}{%
{\protect \APACyear {2022}}%
}]{%
liu2022lofar}
\APACinsertmetastar {%
liu2022lofar}%
\begin{APACrefauthors}%
Liu, N\BPBI Y.%
, Scholten, O.%
, Hare, B\BPBI M.%
, Dwyer, J\BPBI R.%
, Sterpka, C\BPBI F.%
, Kolma{\v{s}}ov{\'a}, I.%
\BCBL {}\ \BBA {} Santol{\'\i}k, O.%
\end{APACrefauthors}%
\unskip\
\newblock
\APACrefYearMonthDay{2022}{}{}.
\newblock
{\BBOQ}\APACrefatitle {LOFAR observations of lightning initial breakdown
  pulses} {Lofar observations of lightning initial breakdown pulses}.{\BBCQ}
\newblock
\APACjournalVolNumPages{Geophysical Research Letters}{49}{6}{e2022GL098073}.
\PrintBackRefs{\CurrentBib}

\bibitem [\protect \citeauthoryear {%
Luque%
}{%
Luque%
}{%
{\protect \APACyear {2017}}%
}]{%
luque2017radio}
\APACinsertmetastar {%
luque2017radio}%
\begin{APACrefauthors}%
Luque, A.%
\end{APACrefauthors}%
\unskip\
\newblock
\APACrefYearMonthDay{2017}{}{}.
\newblock
{\BBOQ}\APACrefatitle {Radio frequency electromagnetic radiation from streamer
  collisions} {Radio frequency electromagnetic radiation from streamer
  collisions}.{\BBCQ}
\newblock
\APACjournalVolNumPages{Journal of Geophysical Research:
  Atmospheres}{122}{19}{10--497}.
\PrintBackRefs{\CurrentBib}

\bibitem [\protect \citeauthoryear {%
Malla%
, Guo%
, Nijdam%
\BCBL {}\ \BBA {} Teunissen%
}{%
Malla%
\ \protect \BOthers {.}}{%
{\protect \APACyear {2024}}%
}]{%
zenodo-radio}
\APACinsertmetastar {%
zenodo-radio}%
\begin{APACrefauthors}%
Malla, H.%
, Guo, Y.%
, Nijdam, S.%
\BCBL {}\ \BBA {} Teunissen, J.%
\end{APACrefauthors}%
\unskip\
\newblock
\APACrefYearMonthDay{2024}{07}{}.
\newblock
\APACrefbtitle {Data files related to radio emission from streamer discharges
  [Data set].} {Data files related to radio emission from streamer discharges
  [data set].}
\newblock
\APACaddressPublisher{}{Zenodo}.
\newblock
\begin{APACrefURL}
  [{2024-07-26}]\url{https://zenodo.org/doi/10.5281/zenodo.10977532}
  \end{APACrefURL}
\PrintBackRefs{\CurrentBib}

\bibitem [\protect \citeauthoryear {%
Malla%
, Martinez%
, Ebert%
\BCBL {}\ \BBA {} Teunissen%
}{%
Malla%
\ \protect \BOthers {.}}{%
{\protect \APACyear {2023}}%
}]{%
malla2023double}
\APACinsertmetastar {%
malla2023double}%
\begin{APACrefauthors}%
Malla, H.%
, Martinez, A.%
, Ebert, U.%
\BCBL {}\ \BBA {} Teunissen, J.%
\end{APACrefauthors}%
\unskip\
\newblock
\APACrefYearMonthDay{2023}{}{}.
\newblock
{\BBOQ}\APACrefatitle {Double-pulse streamer simulations for varying interpulse
  times in air} {Double-pulse streamer simulations for varying interpulse times
  in air}.{\BBCQ}
\newblock
\APACjournalVolNumPages{Plasma Sources Science and Technology}{32}{9}{095006}.
\PrintBackRefs{\CurrentBib}

\bibitem [\protect \citeauthoryear {%
Nijdam%
, Geurts%
, Van~Veldhuizen%
\BCBL {}\ \BBA {} Ebert%
}{%
Nijdam%
\ \protect \BOthers {.}}{%
{\protect \APACyear {2009}}%
}]{%
Nijdam_2009a}
\APACinsertmetastar {%
Nijdam_2009a}%
\begin{APACrefauthors}%
Nijdam, S.%
, Geurts, C\BPBI G\BPBI C.%
, Van~Veldhuizen, E\BPBI M.%
\BCBL {}\ \BBA {} Ebert, U.%
\end{APACrefauthors}%
\unskip\
\newblock
\APACrefYearMonthDay{2009}{{\APACmonth{02}}}{}.
\newblock
{\BBOQ}\APACrefatitle {Reconnection and Merging of Positive Streamers in Air}
  {Reconnection and merging of positive streamers in air}.{\BBCQ}
\newblock
\APACjournalVolNumPages{Journal of Physics D: Applied Physics}{42}{4}{045201}.
\newblock
\begin{APACrefDOI} \doi{10.1088/0022-3727/42/4/045201} \end{APACrefDOI}
\PrintBackRefs{\CurrentBib}

\bibitem [\protect \citeauthoryear {%
Nijdam%
, Teunissen%
\BCBL {}\ \BBA {} Ebert%
}{%
Nijdam%
\ \protect \BOthers {.}}{%
{\protect \APACyear {2020}}%
}]{%
nijdam_physics_2020}
\APACinsertmetastar {%
nijdam_physics_2020}%
\begin{APACrefauthors}%
Nijdam, S.%
, Teunissen, J.%
\BCBL {}\ \BBA {} Ebert, U.%
\end{APACrefauthors}%
\unskip\
\newblock
\APACrefYearMonthDay{2020}{{\APACmonth{11}}}{}.
\newblock
{\BBOQ}\APACrefatitle {The physics of streamer discharge phenomena} {The
  physics of streamer discharge phenomena}.{\BBCQ}
\newblock
\APACjournalVolNumPages{Plasma Sources Sci. Technol.}{29}{10}{103001}.
\newblock
\begin{APACrefURL} [{2021-11-22}]\url{https://doi.org/10.1088/1361-6595/abaa05}
  \end{APACrefURL}
\newblock
\APACrefnote{Publisher: IOP Publishing}
\newblock
\begin{APACrefDOI} \doi{10.1088/1361-6595/abaa05} \end{APACrefDOI}
\PrintBackRefs{\CurrentBib}

\bibitem [\protect \citeauthoryear {%
Nijdam%
, Teunissen%
, Takahashi%
\BCBL {}\ \BBA {} Ebert%
}{%
Nijdam%
\ \protect \BOthers {.}}{%
{\protect \APACyear {2016}}%
}]{%
Nijdam_2016a}
\APACinsertmetastar {%
Nijdam_2016a}%
\begin{APACrefauthors}%
Nijdam, S.%
, Teunissen, J.%
, Takahashi, E.%
\BCBL {}\ \BBA {} Ebert, U.%
\end{APACrefauthors}%
\unskip\
\newblock
\APACrefYearMonthDay{2016}{{\APACmonth{08}}}{}.
\newblock
{\BBOQ}\APACrefatitle {The Role of Free Electrons in the Guiding of Positive
  Streamers} {The role of free electrons in the guiding of positive
  streamers}.{\BBCQ}
\newblock
\APACjournalVolNumPages{Plasma Sources Science and Technology}{25}{4}{044001}.
\newblock
\begin{APACrefDOI} \doi{10.1088/0963-0252/25/4/044001} \end{APACrefDOI}
\PrintBackRefs{\CurrentBib}

\bibitem [\protect \citeauthoryear {%
Pancheshnyi%
, Nudnova%
\BCBL {}\ \BBA {} Starikovskii%
}{%
Pancheshnyi%
\ \protect \BOthers {.}}{%
{\protect \APACyear {2005}}%
}]{%
pancheshnyi_development_2005}
\APACinsertmetastar {%
pancheshnyi_development_2005}%
\begin{APACrefauthors}%
Pancheshnyi, S.%
, Nudnova, M.%
\BCBL {}\ \BBA {} Starikovskii, A.%
\end{APACrefauthors}%
\unskip\
\newblock
\APACrefYearMonthDay{2005}{{\APACmonth{01}}}{}.
\newblock
{\BBOQ}\APACrefatitle {Development of a cathode-directed streamer discharge in
  air at different pressures: {Experiment} and comparison with direct numerical
  simulation} {Development of a cathode-directed streamer discharge in air at
  different pressures: {Experiment} and comparison with direct numerical
  simulation}.{\BBCQ}
\newblock
\APACjournalVolNumPages{Phys. Rev. E}{71}{1}{016407}.
\newblock
\begin{APACrefURL}
  [{2021-10-26}]\url{https://link.aps.org/doi/10.1103/PhysRevE.71.016407}
  \end{APACrefURL}
\newblock
\APACrefnote{Publisher: American Physical Society}
\newblock
\begin{APACrefDOI} \doi{10.1103/PhysRevE.71.016407} \end{APACrefDOI}
\PrintBackRefs{\CurrentBib}

\bibitem [\protect \citeauthoryear {%
Parkevich%
, Khirianova%
\BCBL {}\ \protect \BOthers {.}}{%
Parkevich%
, Khirianova%
\BCBL {}\ \protect \BOthers {.}}{%
{\protect \APACyear {2022}}%
}]{%
parkevich2022electromagnetic}
\APACinsertmetastar {%
parkevich2022electromagnetic}%
\begin{APACrefauthors}%
Parkevich, E.%
, Khirianova, A.%
, Khirianov, T.%
, Baidin, I.%
, Shpakov, K.%
, Rodionov, A.%
\BDBL {}others%
\end{APACrefauthors}%
\unskip\
\newblock
\APACrefYearMonthDay{2022}{}{}.
\newblock
{\BBOQ}\APACrefatitle {Electromagnetic emissions in the MHz and GHz frequency
  ranges driven by the streamer formation processes} {Electromagnetic emissions
  in the mhz and ghz frequency ranges driven by the streamer formation
  processes}.{\BBCQ}
\newblock
\APACjournalVolNumPages{Physical Review E}{106}{4}{045210}.
\PrintBackRefs{\CurrentBib}

\bibitem [\protect \citeauthoryear {%
Parkevich%
, Shpakov%
\BCBL {}\ \protect \BOthers {.}}{%
Parkevich%
, Shpakov%
\BCBL {}\ \protect \BOthers {.}}{%
{\protect \APACyear {2022}}%
}]{%
parkevich2022streamer}
\APACinsertmetastar {%
parkevich2022streamer}%
\begin{APACrefauthors}%
Parkevich, E.%
, Shpakov, K.%
, Baidin, I.%
, Rodionov, A.%
, Khirianova, A.%
, Khirianov, T.%
\BDBL {}others%
\end{APACrefauthors}%
\unskip\
\newblock
\APACrefYearMonthDay{2022}{}{}.
\newblock
{\BBOQ}\APACrefatitle {Streamer formation processes trigger intense x-ray and
  high-frequency radio emissions in a high-voltage discharge} {Streamer
  formation processes trigger intense x-ray and high-frequency radio emissions
  in a high-voltage discharge}.{\BBCQ}
\newblock
\APACjournalVolNumPages{Physical Review E}{105}{5}{L053201}.
\PrintBackRefs{\CurrentBib}

\bibitem [\protect \citeauthoryear {%
Petersen%
\ \BBA {} Beasley%
}{%
Petersen%
\ \BBA {} Beasley%
}{%
{\protect \APACyear {2014}}%
}]{%
petersen2014microwave}
\APACinsertmetastar {%
petersen2014microwave}%
\begin{APACrefauthors}%
Petersen, D.%
\BCBT {}\ \BBA {} Beasley, W.%
\end{APACrefauthors}%
\unskip\
\newblock
\APACrefYearMonthDay{2014}{}{}.
\newblock
{\BBOQ}\APACrefatitle {Microwave radio emissions of negative cloud-to-ground
  lightning flashes} {Microwave radio emissions of negative cloud-to-ground
  lightning flashes}.{\BBCQ}
\newblock
\APACjournalVolNumPages{Atmospheric research}{135}{}{314--321}.
\PrintBackRefs{\CurrentBib}

\bibitem [\protect \citeauthoryear {%
\APACcitebtitle {{Phelps} database, www.lxcat.net, retrieved on August 19,
  2021.}}{%
\APACcitebtitle {{Phelps} database, www.lxcat.net, retrieved on August 19,
  2021.}}{%
{\protect \APACyear {{\protect \bibnodate {}}}}%
}]{%
Phelps_database}
\APACinsertmetastar {%
Phelps_database}%
\APACrefbtitle {{Phelps} database, www.lxcat.net, retrieved on August 19,
  2021.} {{Phelps} database, www.lxcat.net, retrieved on august 19, 2021.}
\newblock
\APACrefYearMonthDay{{\protect \bibnodate {}}}{}{}.
\PrintBackRefs{\CurrentBib}

\bibitem [\protect \citeauthoryear {%
Pu%
, Cummer%
\BCBL {}\ \BBA {} Liu%
}{%
Pu%
\ \protect \BOthers {.}}{%
{\protect \APACyear {2021}}%
}]{%
Pu_2021}
\APACinsertmetastar {%
Pu_2021}%
\begin{APACrefauthors}%
Pu, Y.%
, Cummer, S\BPBI A.%
\BCBL {}\ \BBA {} Liu, N.%
\end{APACrefauthors}%
\unskip\
\newblock
\APACrefYearMonthDay{2021}{{\APACmonth{06}}}{}.
\newblock
{\BBOQ}\APACrefatitle {{{VHF Radio Spectrum}} of a {{Positive Leader}} and
  {{Implications}} for {{Electric Fields}}} {{{VHF Radio Spectrum}} of a
  {{Positive Leader}} and {{Implications}} for {{Electric Fields}}}.{\BBCQ}
\newblock
\APACjournalVolNumPages{Geophysical Research Letters}{48}{11}{e2021GL093145}.
\newblock
\begin{APACrefDOI} \doi{10.1029/2021GL093145} \end{APACrefDOI}
\PrintBackRefs{\CurrentBib}

\bibitem [\protect \citeauthoryear {%
Pu%
, Liu%
\BCBL {}\ \BBA {} Cummer%
}{%
Pu%
\ \protect \BOthers {.}}{%
{\protect \APACyear {2022}}%
}]{%
Pu_2022}
\APACinsertmetastar {%
Pu_2022}%
\begin{APACrefauthors}%
Pu, Y.%
, Liu, N.%
\BCBL {}\ \BBA {} Cummer, S\BPBI A.%
\end{APACrefauthors}%
\unskip\
\newblock
\APACrefYearMonthDay{2022}{{\APACmonth{03}}}{}.
\newblock
{\BBOQ}\APACrefatitle {Quantification of {{Electric Fields}} in {{Fast
  Breakdown During Lightning Initiation From VHF}}-{{UHF Power Spectra}}}
  {Quantification of {{Electric Fields}} in {{Fast Breakdown During Lightning
  Initiation From VHF}}-{{UHF Power Spectra}}}.{\BBCQ}
\newblock
\APACjournalVolNumPages{Geophysical Research Letters}{49}{5}{e2021GL097374}.
\newblock
\begin{APACrefDOI} \doi{10.1029/2021GL097374} \end{APACrefDOI}
\PrintBackRefs{\CurrentBib}

\bibitem [\protect \citeauthoryear {%
Qin%
, Celestin%
\BCBL {}\ \BBA {} Pasko%
}{%
Qin%
\ \protect \BOthers {.}}{%
{\protect \APACyear {2012}}%
}]{%
Qin_2012b}
\APACinsertmetastar {%
Qin_2012b}%
\begin{APACrefauthors}%
Qin, J.%
, Celestin, S.%
\BCBL {}\ \BBA {} Pasko, V\BPBI P.%
\end{APACrefauthors}%
\unskip\
\newblock
\APACrefYearMonthDay{2012}{{\APACmonth{11}}}{}.
\newblock
{\BBOQ}\APACrefatitle {Low Frequency Electromagnetic Radiation from Sprite
  Streamers} {Low frequency electromagnetic radiation from sprite
  streamers}.{\BBCQ}
\newblock
\APACjournalVolNumPages{Geophysical Research Letters}{39}{22}{2012GL053991}.
\newblock
\begin{APACrefDOI} \doi{10.1029/2012GL053991} \end{APACrefDOI}
\PrintBackRefs{\CurrentBib}

\bibitem [\protect \citeauthoryear {%
Rakov%
\ \BBA {} Uman%
}{%
Rakov%
\ \BBA {} Uman%
}{%
{\protect \APACyear {2003}}%
}]{%
rakov2003lightning}
\APACinsertmetastar {%
rakov2003lightning}%
\begin{APACrefauthors}%
Rakov, V\BPBI A.%
\BCBT {}\ \BBA {} Uman, M\BPBI A.%
\end{APACrefauthors}%
\unskip\
\newblock
\APACrefYear{2003}.
\newblock
\APACrefbtitle {Lightning: physics and effects} {Lightning: physics and
  effects}.
\newblock
\APACaddressPublisher{}{Cambridge university press}.
\PrintBackRefs{\CurrentBib}

\bibitem [\protect \citeauthoryear {%
Scholten%
\ \protect \BOthers {.}}{%
Scholten%
\ \protect \BOthers {.}}{%
{\protect \APACyear {2023}}%
}]{%
Scholten_2023}
\APACinsertmetastar {%
Scholten_2023}%
\begin{APACrefauthors}%
Scholten, O.%
, Hare, B\BPBI M.%
, Dwyer, J.%
, Liu, N.%
, Sterpka, C.%
, Assink, J.%
\BDBL {}Veen, S\BPBI T.%
\end{APACrefauthors}%
\unskip\
\newblock
\APACrefYearMonthDay{2023}{{\APACmonth{04}}}{}.
\newblock
{\BBOQ}\APACrefatitle {Small-{{Scale Discharges Observed Near}} the {{Top}} of
  a {{Thunderstorm}}} {Small-{{Scale Discharges Observed Near}} the {{Top}} of
  a {{Thunderstorm}}}.{\BBCQ}
\newblock
\APACjournalVolNumPages{Geophysical Research Letters}{50}{8}{e2022GL101304}.
\newblock
\begin{APACrefDOI} \doi{10.1029/2022GL101304} \end{APACrefDOI}
\PrintBackRefs{\CurrentBib}

\bibitem [\protect \citeauthoryear {%
Sejdic%
, Djurovic%
\BCBL {}\ \BBA {} Stankovic%
}{%
Sejdic%
\ \protect \BOthers {.}}{%
{\protect \APACyear {2008}}%
}]{%
sejdic2008quantitative}
\APACinsertmetastar {%
sejdic2008quantitative}%
\begin{APACrefauthors}%
Sejdic, E.%
, Djurovic, I.%
\BCBL {}\ \BBA {} Stankovic, L.%
\end{APACrefauthors}%
\unskip\
\newblock
\APACrefYearMonthDay{2008}{}{}.
\newblock
{\BBOQ}\APACrefatitle {Quantitative performance analysis of scaleogram as
  instantaneous frequency estimator} {Quantitative performance analysis of
  scaleogram as instantaneous frequency estimator}.{\BBCQ}
\newblock
\APACjournalVolNumPages{IEEE Transactions on Signal
  Processing}{56}{8}{3837--3845}.
\PrintBackRefs{\CurrentBib}

\bibitem [\protect \citeauthoryear {%
Shao%
}{%
Shao%
}{%
{\protect \APACyear {2016}}%
}]{%
Shao_2016}
\APACinsertmetastar {%
Shao_2016}%
\begin{APACrefauthors}%
Shao, X\BHBI M.%
\end{APACrefauthors}%
\unskip\
\newblock
\APACrefYearMonthDay{2016}{{\APACmonth{04}}}{}.
\newblock
{\BBOQ}\APACrefatitle {Generalization of the Lightning Electromagnetic
  Equations of {{Uman}}, {{McLain}}, and {{Krider}} Based on {{Jefimenko}}
  Equations} {Generalization of the lightning electromagnetic equations of
  {{Uman}}, {{McLain}}, and {{Krider}} based on {{Jefimenko}}
  equations}.{\BBCQ}
\newblock
\APACjournalVolNumPages{Journal of Geophysical Research:
  Atmospheres}{121}{7}{3363--3371}.
\newblock
\begin{APACrefDOI} \doi{10.1002/2015JD024717} \end{APACrefDOI}
\PrintBackRefs{\CurrentBib}

\bibitem [\protect \citeauthoryear {%
Shi%
, Liu%
, Dwyer%
\BCBL {}\ \BBA {} Ihaddadene%
}{%
Shi%
\ \protect \BOthers {.}}{%
{\protect \APACyear {2019}}%
}]{%
shi2019vhf}
\APACinsertmetastar {%
shi2019vhf}%
\begin{APACrefauthors}%
Shi, F.%
, Liu, N.%
, Dwyer, J\BPBI R.%
\BCBL {}\ \BBA {} Ihaddadene, K\BPBI M.%
\end{APACrefauthors}%
\unskip\
\newblock
\APACrefYearMonthDay{2019}{}{}.
\newblock
{\BBOQ}\APACrefatitle {VHF and UHF electromagnetic radiation produced by
  streamers in lightning} {Vhf and uhf electromagnetic radiation produced by
  streamers in lightning}.{\BBCQ}
\newblock
\APACjournalVolNumPages{Geophysical Research Letters}{46}{1}{443--451}.
\PrintBackRefs{\CurrentBib}

\bibitem [\protect \citeauthoryear {%
Shi%
, Liu%
\BCBL {}\ \BBA {} Rassoul%
}{%
Shi%
\ \protect \BOthers {.}}{%
{\protect \APACyear {2016}}%
}]{%
shi2016properties}
\APACinsertmetastar {%
shi2016properties}%
\begin{APACrefauthors}%
Shi, F.%
, Liu, N.%
\BCBL {}\ \BBA {} Rassoul, H\BPBI K.%
\end{APACrefauthors}%
\unskip\
\newblock
\APACrefYearMonthDay{2016}{}{}.
\newblock
{\BBOQ}\APACrefatitle {Properties of relatively long streamers initiated from
  an isolated hydrometeor} {Properties of relatively long streamers initiated
  from an isolated hydrometeor}.{\BBCQ}
\newblock
\APACjournalVolNumPages{Journal of Geophysical Research:
  Atmospheres}{121}{12}{7284--7295}.
\PrintBackRefs{\CurrentBib}

\bibitem [\protect \citeauthoryear {%
Stephane%
}{%
Stephane%
}{%
{\protect \APACyear {1999}}%
}]{%
stephane1999wavelet}
\APACinsertmetastar {%
stephane1999wavelet}%
\begin{APACrefauthors}%
Stephane, M.%
\end{APACrefauthors}%
\unskip\
\newblock
\APACrefYearMonthDay{1999}{}{}.
\newblock
\APACrefbtitle {A wavelet tour of signal processing.} {A wavelet tour of signal
  processing.}
\newblock
\APACaddressPublisher{}{Elsevier}.
\PrintBackRefs{\CurrentBib}

\bibitem [\protect \citeauthoryear {%
Teunissen%
\ \BBA {} Ebert%
}{%
Teunissen%
\ \BBA {} Ebert%
}{%
{\protect \APACyear {2017}}%
}]{%
teunissen_simulating_2017}
\APACinsertmetastar {%
teunissen_simulating_2017}%
\begin{APACrefauthors}%
Teunissen, J.%
\BCBT {}\ \BBA {} Ebert, U.%
\end{APACrefauthors}%
\unskip\
\newblock
\APACrefYearMonthDay{2017}{oct}{}.
\newblock
{\BBOQ}\APACrefatitle {Simulating streamer discharges in 3D with the parallel
  adaptive Afivo framework} {Simulating streamer discharges in 3d with the
  parallel adaptive afivo framework}.{\BBCQ}
\newblock
\APACjournalVolNumPages{Journal of Physics D: Applied Physics}{50}{47}{474001}.
\newblock
\begin{APACrefURL} \url{https://doi.org/10.1088/1361-6463/aa8faf}
  \end{APACrefURL}
\newblock
\begin{APACrefDOI} \doi{10.1088/1361-6463/aa8faf} \end{APACrefDOI}
\PrintBackRefs{\CurrentBib}

\bibitem [\protect \citeauthoryear {%
Teunissen%
\ \BBA {} Ebert%
}{%
Teunissen%
\ \BBA {} Ebert%
}{%
{\protect \APACyear {2018}}%
}]{%
Teunissen_2018_afivo}
\APACinsertmetastar {%
Teunissen_2018_afivo}%
\begin{APACrefauthors}%
Teunissen, J.%
\BCBT {}\ \BBA {} Ebert, U.%
\end{APACrefauthors}%
\unskip\
\newblock
\APACrefYearMonthDay{2018}{{\APACmonth{12}}}{}.
\newblock
{\BBOQ}\APACrefatitle {Afivo: A Framework for Quadtree/Octree {{AMR}} with
  Shared-Memory Parallelization and Geometric Multigrid Methods} {Afivo: A
  framework for quadtree/octree {{AMR}} with shared-memory parallelization and
  geometric multigrid methods}.{\BBCQ}
\newblock
\APACjournalVolNumPages{Computer Physics Communications}{233}{}{156--166}.
\newblock
\begin{APACrefDOI} \doi{10.1016/j.cpc.2018.06.018} \end{APACrefDOI}
\PrintBackRefs{\CurrentBib}

\bibitem [\protect \citeauthoryear {%
Torrence%
\ \BBA {} Compo%
}{%
Torrence%
\ \BBA {} Compo%
}{%
{\protect \APACyear {1998}}%
}]{%
torrence1998practical}
\APACinsertmetastar {%
torrence1998practical}%
\begin{APACrefauthors}%
Torrence, C.%
\BCBT {}\ \BBA {} Compo, G\BPBI P.%
\end{APACrefauthors}%
\unskip\
\newblock
\APACrefYearMonthDay{1998}{}{}.
\newblock
{\BBOQ}\APACrefatitle {A practical guide to wavelet analysis} {A practical
  guide to wavelet analysis}.{\BBCQ}
\newblock
\APACjournalVolNumPages{Bulletin of the American Meteorological
  society}{79}{1}{61--78}.
\PrintBackRefs{\CurrentBib}

\bibitem [\protect \citeauthoryear {%
Uman%
, McLain%
\BCBL {}\ \BBA {} Krider%
}{%
Uman%
\ \protect \BOthers {.}}{%
{\protect \APACyear {1975}}%
}]{%
uman1975electromagnetic}
\APACinsertmetastar {%
uman1975electromagnetic}%
\begin{APACrefauthors}%
Uman, M\BPBI A.%
, McLain, D\BPBI K.%
\BCBL {}\ \BBA {} Krider, E\BPBI P.%
\end{APACrefauthors}%
\unskip\
\newblock
\APACrefYearMonthDay{1975}{}{}.
\newblock
{\BBOQ}\APACrefatitle {The electromagnetic radiation from a finite antenna}
  {The electromagnetic radiation from a finite antenna}.{\BBCQ}
\newblock
\APACjournalVolNumPages{American Journal of Physics}{43}{1}{33--38}.
\PrintBackRefs{\CurrentBib}

\bibitem [\protect \citeauthoryear {%
Wang%
\ \protect \BOthers {.}}{%
Wang%
\ \protect \BOthers {.}}{%
{\protect \APACyear {2023}}%
}]{%
Wang_2023}
\APACinsertmetastar {%
Wang_2023}%
\begin{APACrefauthors}%
Wang, Z.%
, Dijcks, S.%
, Guo, Y.%
, Van Der~Leegte, M.%
, Sun, A.%
, Ebert, U.%
\BDBL {}Teunissen, J.%
\end{APACrefauthors}%
\unskip\
\newblock
\APACrefYearMonthDay{2023}{{\APACmonth{08}}}{}.
\newblock
{\BBOQ}\APACrefatitle {Quantitative Modeling of Streamer Discharge Branching in
  Air} {Quantitative modeling of streamer discharge branching in air}.{\BBCQ}
\newblock
\APACjournalVolNumPages{Plasma Sources Science and Technology}{32}{8}{085007}.
\newblock
\begin{APACrefDOI} \doi{10.1088/1361-6595/ace9fa} \end{APACrefDOI}
\PrintBackRefs{\CurrentBib}

\bibitem [\protect \citeauthoryear {%
Wang%
, Sun%
\BCBL {}\ \BBA {} Teunissen%
}{%
Wang%
\ \protect \BOthers {.}}{%
{\protect \APACyear {2022}}%
}]{%
Wang_2022}
\APACinsertmetastar {%
Wang_2022}%
\begin{APACrefauthors}%
Wang, Z.%
, Sun, A.%
\BCBL {}\ \BBA {} Teunissen, J.%
\end{APACrefauthors}%
\unskip\
\newblock
\APACrefYearMonthDay{2022}{{\APACmonth{01}}}{}.
\newblock
{\BBOQ}\APACrefatitle {A Comparison of Particle and Fluid Models for Positive
  Streamer Discharges in Air} {A comparison of particle and fluid models for
  positive streamer discharges in air}.{\BBCQ}
\newblock
\APACjournalVolNumPages{Plasma Sources Science and Technology}{31}{1}{015012}.
\newblock
\begin{APACrefDOI} \doi{10.1088/1361-6595/ac417b} \end{APACrefDOI}
\PrintBackRefs{\CurrentBib}

\bibitem [\protect \citeauthoryear {%
{Zheleznyak}%
, {Mnatsakanian}%
\BCBL {}\ \BBA {} {Sizykh}%
}{%
{Zheleznyak}%
\ \protect \BOthers {.}}{%
{\protect \APACyear {1982}}%
}]{%
zheleznyak1982photoi_english}
\APACinsertmetastar {%
zheleznyak1982photoi_english}%
\begin{APACrefauthors}%
{Zheleznyak}, M\BPBI B.%
, {Mnatsakanian}, A\BPBI K.%
\BCBL {}\ \BBA {} {Sizykh}, S\BPBI V.%
\end{APACrefauthors}%
\unskip\
\newblock
\APACrefYearMonthDay{1982}{{\APACmonth{11}}}{}.
\newblock
{\BBOQ}\APACrefatitle {{Photoionization of nitrogen and oxygen mixtures by
  radiation from a gas discharge}} {{Photoionization of nitrogen and oxygen
  mixtures by radiation from a gas discharge}}.{\BBCQ}
\newblock
\APACjournalVolNumPages{High Temperature Science}{20}{3}{357-362}.
\PrintBackRefs{\CurrentBib}

\end{thebibliography}

%
%
%
%
%
\appendix

\section{Numerically solving Jefimenko's equations on an AMR grid}
\label{sec:jefimenko-numerical}

Below we explain how we numerically evaluate the integral of equation~(\ref{eqn:jefimenko_E}) to obtain the radiated electric field.
In our results, we have omitted the first (static) term, which does not correspond to the radiated field and which can be made arbitrarily small by picking a farther away observation location.
There are several factor that make the evaluation of the other two terms non-trivial.
First, strict charge conservation is required, otherwise there can be a large artificial signal from the $\partial_t \rho$ term.
We guarantee this up to the level op machine precision by employing conservative interpolation and coarsening methods in our AMR simulations.
Furthermore, we ensure the discharges stay away from the domain boundaries, so that there is no charge flowing in or out of the domain.

Another challenge is that time derivatives of $\rho$ and $\mathbf{J}$ are required.
To obtain these at time $t_1$, we use central differencing to obtain
\begin{equation}
  \partial_t \rho(\mathbf{r}, t_1) \approx \frac{\rho(\mathbf{r}, t_2) - \rho(\mathbf{r}, t_0)}{2 \Delta t},
  \label{eq:rho-deriv}
\end{equation}
where $\Delta t$ is the (constant) time step between simulation outputs.
This is not enough yet, because the integral requires this derivative at different retarded times, with $t_r = t - \mathcal{R}/c$ having a different value for every grid cell.
We therefore compute a second derivative as
\begin{equation}
  \partial_t^2 \rho(\mathbf{r}, t_1) \approx \frac{\rho(\mathbf{r}, t_2) - 2 \rho(\mathbf{r}, t_1) +  \rho(\mathbf{r}, t_0)}{\Delta t^2}.
  \label{eq:rho-deriv2}
\end{equation}
We can then account for the variable time retardation by approximating
\begin{equation}
  \partial_t \rho(\mathbf{r}, t_1 + h) \approx \partial_t \rho(\mathbf{r}, t_1) + h \partial_t^2 \rho(\mathbf{r}, t_1).
  \label{eq:rho-approx}
\end{equation}
The same approach is used for each component of the current density $\mathbf{J}$.

Since the AMR grid changes in time, we have to map all the quantities appearing in equations~(\ref{eq:rho-deriv})--~(\ref{eq:rho-approx}) to the same grid.
At first, we used a uniform grid for this purpose, which was limited to $128^3$ or $256^3$ due to memory limitations.
However, this did not lead to satisfactory results, because the motion of the streamer was not captured smoothly on such a grid, leading to artificial signals when streamer's head charge would move from one cell to the next.
We therefore decided to instead map $\rho(\mathbf{r}, t_1)$ and $\rho(\mathbf{r}, t_2)$ (and similar for the components of $\mathbf{J}$) to the AMR grid corresponding to $\rho(\mathbf{r}, t_0)$ in these equations.
This was again done in a conservative way.
Finally, for computational efficiency, we decided to save the simulation output not up to the finest AMR level used in the discharge simulations, but up to a resolution of about $0.15 \, \textrm{mm}$.

\section{The far field approximation for electromagnetic waves}
\label{sec:far-field}

Equation (\ref{eq:E=cB}) quotes the approximate relation 
$E_\mathrm{rad} = c \, B_\mathrm{rad}$
between the size of magnetic and electric field in an electromagnetic wave far from the source. Here we discuss the range of validity of this approximation.

The textbook by \citeA{griffiths2005introduction} (as well as many other sources) contains the derivation of the relation 
\begin{equation} \label{eq:C1}
    {\bf B} = \frac{\hat{\bf k}\times {\bf E}}c
\end{equation}
for a monochromatic plane wave ${\bf E}({\bf r},t) = {\bf E}\;e^{i({\bf k\cdot r}-\omega t)}$ in vacuum. Here {\bf k} is the wave vector, and $\hat{\bf k} = {\bf k}/|{\bf k}|$ is the unit vector parallel to the wave vector.
${\bf E}$ is orthogonal to {\bf k} by construction, so ${\bf k}\cdot {\bf E}=0$ and 
$|\hat{\bf k}\times {\bf E}| =  |{\bf E}|$, which together with (\ref{eq:C1}) means that
\begin{equation}   \label{eq:mono}
    |{\bf E}| = c |{\bf B}|
\end{equation}
in this particular case. But is this relation also valid more generally, in particular for an arbitrary superposition of Fourier modes?

An answer can be found by superposing two Fourier modes with different wave vectors ${\bf k}_1$ and ${\bf k}_2$, so that ${\bf E}= {\bf E}_{1} + {\bf E}_{2}$ and ${\bf B}= (\hat{\bf k}_1 \times \mathbf{E}_1 + \hat{\bf k}_2 \times \mathbf{E}_2)/c$.
It then follows that
\begin{equation} 
    |{\bf E}|^2 - c^2 |{\bf B}|^2 = 2 \Big(
    {\bf E}_1\cdot {\bf E}_2 - 
    (\hat{\bf k}_1\times {\bf E}_1) \cdot (\hat{\bf k}_2\times {\bf E}_2)
    \Big),
\end{equation}
which is zero if $\hat{\bf k}_1 = \hat{\bf k}_2$, but not in general.
We can conclude that equation~(\ref{eq:C1})
is applicable to a superposition of planar Fourier modes propagating in the same direction $\hat{\bf k}$, but not to modes propagating at arbitrary directions to each other. The equation can be applied in the far field approximation,
i.e., if the distance to the radiating object is much larger than the object size and the wave length.

Additionally, it can be tested whether magnetic and electric field are still orthogonal
in the superposition of the two modes.
Using the relation (\ref{eq:C1}) for the magnetic fields, the orthogonality relations of the fields in the Fourier modes and standard vector algebra, one finds in a few steps of calculation that
\begin{equation}
    {\bf E}\cdot c{\bf B} = (\hat{\bf k}_1 - \hat{\bf k}_2) \cdot 
    ({\bf E}_{1} \times {\bf E}_{2}).
\end{equation}
This means that electric and magnetic field in the superposition wave are orthogonal to each other, if the wave vectors of the two Fourier modes point in the same direction, $\hat{\bf k}_1 = \hat{\bf k}_2$, but not in general.
This criterion is sufficient for the orthogonality of {\bf E} and {\bf B}, and it is the same as for the validity of equations (\ref{eq:C1}) and (\ref{eq:mono}): the electromagnetic wave is planar, but not necessarily monochromatic.
In other words: the wave can consist of an arbitrary superposition of Fourier modes, if all wave vectors are pointing in the same direction.

\section{\jt{Monte Carlo photoionization}}
\label{sec:mc-photo}

\jt{Below, we briefly describe the Monte Carlo photoionization approach.
For the results presented here, the photoionization source term $S_\mathrm{ph}$ from equation~(\ref{eq:ddt_e}) was upated every two time steps.
If this interval is called $\Delta t_\gamma$, then the algorithm proceeds according to the following steps:
\begin{enumerate}
  \item Initially, set $S_\mathrm{ph}$ to zero.
  \item Determine the expected number of ionizing photons $n_\gamma$ produced in each grid cell during $\Delta t_\gamma$. We use Zheleznyak's model, in which $n_\gamma$ is proportional to the number of electron impact ionization events that took place in the grid cell, see \cite{zheleznyak1982photoi_english} for details.
  \item In each grid cell, use pseudo-random numbers to sample an integer number of photons $N_\gamma$ produced in that cell.
  Each photon has a `weight' of one, i.e., it corresponds to a photon producing a single ionization event.
  \item For each produced photon, sample a random isotropic direction. Next, an absorption distance is sampled according to Zheleznyak's model, as described in~\cite{chanrion2008pic}, and the grid cell in which the photon is absorbed is determined. The photoionization source term $S_\mathrm{ph}$ corresponding to that grid cell is then increased by $(\Delta V \Delta t_\gamma)^{-1}$.
\end{enumerate}

For a more detailed description of implementation of Monte Carlo photoionization, readers are referred to~\citeA{bagheri2019effect} and \citeA{Wang_2023}.}

\end{document}